\documentclass[fleqn,usenatbib]{mnras}

\usepackage[T1]{fontenc}
\usepackage{ae,aecompl}

\usepackage{graphicx}	
\usepackage{amsmath}	
\usepackage{amssymb}	
\usepackage{hyperref}
\usepackage{newtxtext,newtxmath}

\newcommand{\correction}{\color{black}}

\title[Unveiling the Planet Population at Birth]{Unveiling the Planet Population at Birth}

\author[Rogers, J. G. \& Owen, J. E.]{
James G. Rogers\thanks{E-mail: james.rogers14@imperial.ac.uk} and James E. Owen
\\
Astrophysics Group, Department of Physics, Imperial College London, Prince Consort Rd, London, SW7 2AZ, UK\\
}

\date{Accepted XXX. Received YYY; in original form ZZZ}

\pubyear{2020}

\begin{document}
\label{firstpage}
\pagerange{\pageref{firstpage}--\pageref{lastpage}}
\maketitle

\begin{abstract}
The radius distribution of small, close-in exoplanets has recently been shown to be bimodal. The photoevaporation model predicted this bimodality. In the photoevaporation scenario, some planets are completely stripped of their primordial H/He atmospheres, whereas others retain them. Comparisons between the photoevaporation model and observed planetary populations have the power to unveil details of the planet population inaccessible by standard observations, such as the core mass distribution and core composition.  In this work, we present a hierarchical inference analysis on the distribution of close-in exoplanets using forward-models of photoevaporation evolution. We use this model to constrain the planetary distributions for core composition, core mass and initial atmospheric mass fraction. We find that the core-mass distribution is peaked, with a peak-mass of $\sim 4$~M$_\oplus$. The bulk core-composition is consistent with a rock/iron mixture that is ice-poor and ``Earth-like''; the spread in core-composition is found to be narrow ($\lesssim 16\%$ variation in iron-mass fraction at the 2$\sigma$ level) and consistent with zero. This result favours core formation in a water/ice poor environment. We find the majority of planets accreted a H/He envelope with a typical mass fraction of $\sim 4\%$; only a small fraction did not accrete large amounts of H/He and were ``born-rocky''. We find four-times as many super-Earths were formed through photoevaporation, as formed without a large H/He atmosphere. Finally, we find core-accretion theory over-predicts the amount of H/He cores would have accreted by a factor of $\sim 5$, pointing to additional mass-loss mechanisms (e.g. ``boil-off'') or modifications to core-accretion theory. 
\end{abstract}




\section{Introduction} \label{sec:Intro}
A decade since the launch of NASA's \textit{Kepler Space Telescope} \citep{BoruckiKeplerII}, over 4000 extra-solar planets have now been confirmed. Of these, the vast majority are small ($\lesssim$ $4R_\oplus$), low mass ($\lesssim$ $50M_\oplus$) and located close to their host star ($\lesssim$ 100 days) \citep{Howard2010,Batalha2013,Petigura2013,Mullally2015}, demonstrating a stark difference to the planets of our own solar system. As population studies imply $> 30\%$ of GK stars host one or more of these planets \citep{Fressin2013,Silburt2015,Mulders2018,Zhu2018,Zink2019}, understanding their origins is a key challenge in this field. 

One popular planet formation theory that was developed in response to the ubiquity of such planets is the in-situ model \citep[e.g.][]{Hansen2012,Chiang2013}, in which planetary embryos form in the inner disc, close to the location we observe them at today. As a result, the constituents that accreted to build up their cores are the silicate materials that drifted into the inner-disc \citep{Chatterjee2014,Jankovic2018}. One would thus predict a core composition of such planets to be similar to that of Earth. Another approach is the migration model \citep[e.g.][]{Ida2005,Ida2010,Bodenheimer2014,Raymond2014,Bitsch2018b, Raymond2018}, in which planets form further out whilst immersed in a solid enhanced region beyond the water ice-line. The planets then migrate inwards to the orbital period we observe them at today and will therefore have core compositions consistent with one rich in water/ice. In reality, core formation may draw upon physics from both schemes as it is well established that solids must migrate towards the star in the form of pebbles or embryos. What is not clear is which of the models is the main driver of core formation. 

To test the formation models, one can combine transit, radial velocity (RV) and transit timing variation (TTV) data to calculate the bulk density \citep{Weiss2014,HaddenLithwick2014,Dressing2015,JontofHutter2016}. Studies have shown that many planets $\lesssim 2 R_\oplus$ have a density consistent with that of Earth \citep[e.g.][]{Dressing2015,Dorn2019}, whilst larger planets have lower bulk densities, consistent with a rocky core surrounded by H/He atmospheres \citep[e.g.][]{Rogers2015,WolfgangLopez2015}. Alternatively, these reduced densities are also consistent with `water-worlds' \citep[e.g.][]{Valencia2007,Zeng2019}. Clearly, bulk densities alone are not capable of differentiating between the two planet formation models \citep[e.g.][]{RogersSeager2010}.

The degeneracy between internal compositions illuminates a larger problem in understanding planet formation. It demonstrates that standard exoplanet survey data provides only highly correlated information on three important quantities required to place constraints on planet formation models: the core mass distribution, the H/He atmospheric mass fraction distribution and, as discussed, the core density distribution. However, progress can be made if we exploit an evolutionary process, namely EUV/X-ray photoevaporation \citep{Lammer2003,Baraffe2004,MurrayClay2009,Owen2013,Jin2014,LopezFortney2013,ChenRogers2016}, that sculpts the exoplanet population and allows one to `rewind the clock' of planet evolution and reveal the distributions of interest.

The photoevaporation model gained success in predicting one of the most intriguing features of planet demographics; a bimodal distribution in the sizes of small, close-in exoplanets \citep{Owen2013,LopezFortney2013}. In this model, many planets receive an integrated high-energy stellar luminosity in the first few 100 Myr \citep{Jackson2012} that is comparable to the binding energy of their atmospheres \citep{Lammer2003,Lecavelier2007,Davis2009}, which can result in significant atmospheric mass loss. Depending on their initial conditions, close-in planets either maintain an extended H/He atmosphere, or have their H/He atmosphere completely stripped, leaving a bare core \citep{Owen2017,Owen2019}. It was suggested that this dichotomy in atmospheric evolution would produce an abundance of planets detected at their core radius (i.e. a sample of bare cores), and another detected at approximately double their core radius (i.e. cores with extended H/He atmosphere). Indeed, in \cite{Fulton2017,VanEylen2018}, two peaks were observed at $\sim1.3 R_{\oplus}$ and $\sim 2.4 R_{\oplus}$, labelled as `super-Earths' and `sub-Neptunes' respectively. In addition to this observation, confirmation of the photoevaporation model has come from transit spectroscopy of close-in exoplanets in which outflowing atmospheric gas causes increased absorption of atomic lines such as Lyman-$\alpha$ or He-I \citep{VidalMadjar2003,Kulow2014,Ehrenreich2015,Spake2018}. 

In \cite{OwenMorton2016}, photoevaporation was used to break internal structure degeneracies of \textit{Kepler}-36b and c. It was shown that by incorporating an evolutionary model, constraints could be placed on the core mass, atmospheric mass fraction and core density of the planets, the latter of which was consistent with an Earth-like composition for both planets. This work showed the power of incorporating an evolutionary model: not all compositions consistent with a planet's measured properties today are consistent with it's evolutionary history. \citet{Kubyshkina2019b,Kubyshkina2019a} have applied this approach to a number of other planetary systems, even constraining the activity evolution of the star. However, this kind of analysis is only applicable to planets with measured masses; the vast majority of which the observed exoplanet population do not possess.

However, the power of the exoplanet statistics is in number, therefore information contained in the radius distribution alone can be used to learn about composition.  \citet{Owen2017}, compared evolutionary models including photoevaporation to the observed bimodal distribution in planet sizes from the California-Kepler Survey (CKS) \citep{Fulton2017}. This involved choosing a core mass distribution, initial atmospheric mass fraction distribution, and a core composition to tune the final evolved population to the data. Although phenomenological in nature, this work provided further evidence of an Earth-like composition for close-in exoplanets, as well as a typical core mass of a few Earth masses. In \cite{Wu2019}, further work was done to constrain the initial atmospheric mass fraction and core mass distributions by fitting the exoplanet radius histogram, concluding the core-mass distribution was peaked and Earth-like in composition. These works clearly demonstrated that the data could be neatly explained via the photoevaporation process and that the bimodal distribution of super-Earths and sub-Neptunes arose from the same underlying populations of planets. It thus did not depend upon more complex formation processes, such as those which require two populations of planets to fit the data. On the other hand, these works lacked a rigorous statistical inference methodology. They were also restrictive in their distribution functional form and assumed an identical core density for all planets. In this work, we use an evolutionary model which includes photoevaporation to robustly fit the exoplanet radius and period distributions simultaneously. In doing so we are able to ``wind back the clock'' and unveil the properties of the planet population at birth. 

In Section \ref{sec:Method} we present the hierarchical inference model required to place constraints on the distributions of interest. This involves the choice of data to compare with, as well as methodology for incorporating the detection efficiency and measurement uncertainty. In Section \ref{sec:Results}, we present results from the inference model and discuss their implications on planet formation models in Section \ref{sec:Discussion}.

\section{Method} \label{sec:Method}
 This study invokes the photoevaporation process to infer properties of {\it Kepler} planets that are undetectable from standard survey techniques. Specifically, we wish to infer the following population demographics: core-mass distribution, $f(M_\text{core})$, initial atmospheric mass fraction distribution\footnote{I.e. the atmospheric mass fraction distribution at the start of the photoevaporation process, typically after protoplanetary disc dispersal.}, $f(X_\text{atm}^\text{init} \equiv M_\text{atm} / M_\text{core})$, and core density distribution, $f(\rho_\text{core})$. Whilst directly computing exact synthetic populations of planets from these distributions is impossible, one can sample from the distributions and thus synthesise planet populations and compare with real data. To this end, we forward model an ensemble of planets, each with a separate core mass, ($M_\text{core}$), initial atmospheric mass fraction ($X_\text{atm}^\text{init}$), and core density ($\rho_\text{core}$) drawn from the above distributions. By then performing a synthetic transit survey on the evolved planets and comparing to real {\it Kepler} data, we determine how well the population distributions can reproduce the data. One crucial philosophy we adopt in this work is that any inference of demographic properties should be done by comparing synthetic measurements with real observations, as opposed to completeness corrected data compared directly to the models. In this manner, we accurately incorporate the noise and biases inherent in the data into our model and hence have a better handle on the underlying property of question. Our adopted method is thus akin to Bayesian Hierarchical Model\footnote{Although it is technically not a Bayesian Hierarchical Model: this is because our model of the underlying exoplanet population cannot be computed explicitly, and is thus generated by random sampling (as discussed in the following sections).}.
 
 An important caveat is that all results from this work will be conditioned on the fact that photoevaporation is the leading driver in exoplanet evolution. Whilst this model is successful in explaining a cause for the bimodal distribution in planet radii, other models have been proposed. The leading alternative is the core-powered mass-loss model \citep{Ginzburg2018, Gupta2019, Gupta2020}, which draws upon the accretion luminosity from a planet's core in order to strip it's atmosphere and produce the observed sparsity of planets at $\sim 1.8R_\oplus$. Further, \citet{Zeng2019} has also proposed that the bimodal radius distribution is created through two independent formation mechanisms. The impact of adopting a different evolutionary model is discussed in Section \ref{sec:Discussion}.
 
\begin{figure}  
	\includegraphics[width=\columnwidth]{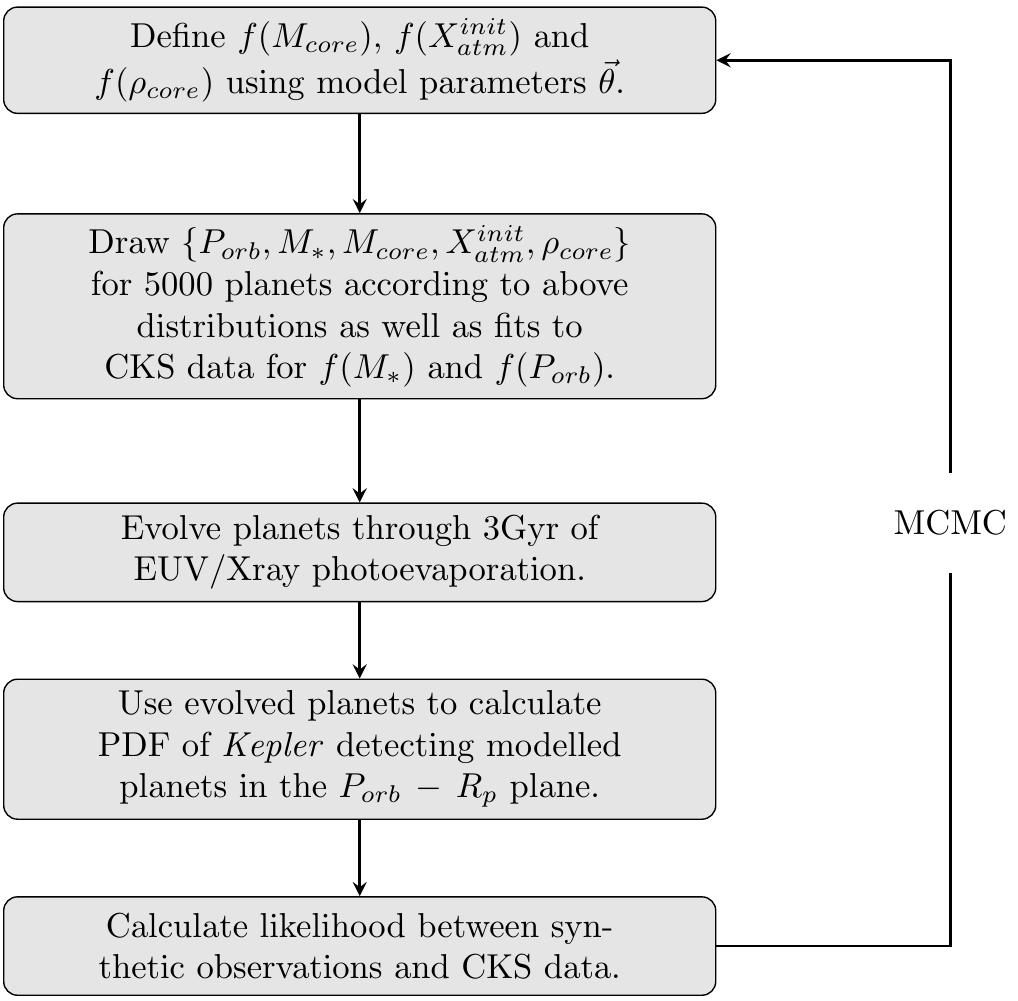}
    \caption{Flowchart to demonstrate the hierarchical inference process. Each MCMC iteration involves the construction of planetary distributions from model parameters, from which 5000 planets are evolved through EUV/X-ray photoevaporation and used to calculate a PDF for planet detection in the orbital period - radius plane. Finally, a likelihood is calculated between the synthetic data and the CKS data set. Each of these five-steps are described separately in Sections \ref{sec:Distributions} - \ref{sec:Likelihood}.}
    \label{fig:FlowChart}
\end{figure}

 The natural choice for the data set comes from the California Kepler Survey (CKS, \citealt{CKSI-Petigura2017,Johson-CKSII2017}) - the sample is large, with well defined cuts and is of high purity. Additionally, the derived planetary radii are sufficiently accurate to resolve a bimodality in planet sizes between $\sim 1 - 4 R_\oplus$ \citep{Fulton2017}. We discuss how additional data sets, particularly RV/TTV mass measurements could be incorporated into the model in Section \ref{sec:Discussion}. One advance over the study of \cite{Wu2019}, which only used the planetary radius distribution in their inference method, is that we choose to compare our models in the orbital period - radius plane, which gives us far greater leverage on the data. Photoevaporation is after all a period dependent process. Figure \ref{fig:FlowChart} shows a schematic outline of the inference problem, which is split into five sections; (1) Constructing the distributions, (2) Drawing the planets, (3) Evolving the planets, (4) Observing the planets and (5) Calculating a likelihood. We will now follow the order of this outline and describe each of these individual processes in the following subsections.
 
\subsection{Constructing the Distributions} \label{sec:Distributions}
In order to forward model exoplanets in this framework, we require distributions for core mass, initial atmospheric mass fraction and bulk core density. For the latter, we choose to quantify this by adopting the approach from \cite{OwenMorton2016}, whereby core density is interpreted by a single parameter $\tilde{\rho} \in [-1,1]$, which tracks the linear fraction of ice, rock and iron as used in mass-radius relationships from \cite{Fortney2007}. A composition of $\tilde{\rho} \leq 0$ signifies a ice-rock mixture, with $\tilde{\rho} = -1$ implying a $100\%$ ice core, ranging to $\tilde{\rho} = 0$ implying a $100\%$ rocky core. Similarly, $\tilde{\rho} \geq 0$ relates to a rock-iron mixture, with $\tilde{\rho}=1$ resulting in a $100\%$ iron core\footnote{We emphasise that the actual quantity we are constraining is the bulk core density, yet we choose to interpret this density in terms of a composition according to a specific mass-radius relationship.}. In this parameterisation the Earth has a value of $1/3$ implying a 1/3 iron, 2/3 rock mass fraction in the \citet{Fortney2007} mass-radius relations. This parameterisation allows us to use a single parameter to specify the core density. In Section \ref{sec:Discussion}, we expand this parameterisation and constrain the full water/rock/iron distribution of the cores. 

Choosing a functional form for these planet properties, namely core mass, initial atmospheric mass fraction and core density is a challenge as there is insufficient observational or theoretical information to make an informed choice for any distribution. In \cite{Wu2019}, the functional forms for core mass and initial atmospheric mass fraction are log-Gaussians, and all planets share the same mean core composition. In the first part of this work, we aim perform a similar analysis to  \cite{Wu2019}, but comparing to the data in both radius and period, rather than radius alone. Thus the distributions for $M_\text{core}$ and $X_\text{atm}^\text{init}$ are set to be log-Gaussians:
\begin{equation} \label{eq:logGauss}
    \begin{split}
        \frac{\mathrm{d} N}{\mathrm{d} \log M_\text{core}} & \; \propto \exp \bigg[- \frac{(\log M_\text{core} - \log \mu_M )^2}{2\sigma^2_M}  \bigg], \\
        \frac{\mathrm{d} N}{\mathrm{d} \log X_\text{atm}^\text{init}} & \; \propto \exp \bigg[- \frac{(\log X_\text{atm}^\text{init} - \log \mu_X )^2}{2\sigma^2_X}  \bigg].
    \end{split}
\end{equation}
The model parameters for these distributions are thus the means and standard deviations for each respective function (e.g $\{ \mu_M, \sigma_M, \mu_X, \sigma_X \}$). Thus, with regards to the planet distributions, the model parameters we wish to infer are $\boldsymbol{\theta} = \{ \tilde{\rho}, \mu_M,  \sigma_M, \mu_X, \sigma_X \}$. We label this simulation as \textsc{model I}.

We then go on to relax the constraint of a specific functional form for core mass and initial atmospheric mass fraction to allow these distributions to be completely arbitrary. Despite being a more challenging inference problem, this choice is necessary as one can now infer new and unexplored features in the distributions of interest, e.g. are the distributions truly peaked, are they skewed? To do so, we employ Bernstein polynomials to define non-parametric distributions for $M_\text{core}$ and $X_\text{atm}^\text{init}$. Similar to other expansions, Bernstein polynomials are capable of approximating any well-behaved function and are an attractive choice as they have convenient degree reduction properties that simplify inference problems. More information can be found in \cite{Farouki2012} or \cite{Bo2018} and in Appendix \ref{app:Bernstein}. Thus, in addition to a mean core composition $\tilde{\rho}$, our model parameters are changed to be the Bernstein polynomial coefficients that define the probability density functions for $\log M_{\text{core}}$ and $\log X_\text{atm}^\text{init}$. We assume uniform priors on these coefficients in the domain $[0,1]$, which thus leads to a log-uniform prior on the core mass and atmosphere mass fraction distributions. The relaxed inference problem thus has the following parameters to fit: $N_M$ coefficients for the core mass PDF, $N_X$ coefficients for initial atmospheric mass fraction PDF and one value for the mean core composition. While their is a strong theoretical prejudice to suspect the initial atmospheric mass fraction correlates with core mass, we choose not to implement this coupling here. This correlation can be explored in future work. This simulation is labelled as \textsc{model II}

Our third and final model is similar to \textsc{model II} in that we use Bernstein polynomials to constrain the core mass and initial atmospheric mass fraction distribution. However, unlike \textsc{model II}, we also attempt to fit the core composition distribution with a Gaussian function. As shown in \cite{Owen2017}, the core density of a planet controls the maximum size at which it can be stripped (at a given core mass and orbital period) and therefore strongly controls the location of the radius gap. The effect of allowing a range of core densities in a given population is to `smear' out the radius gap and thus reduce its depth. In order to maintain the observed sparsity of planets $\sim 1.8R_\oplus$ in our model, we already expect the width of this Gaussian function to be very narrow \citep{Owen2017,VanEylen2018}; however it is yet to be quantified. The final simulation, including the Bernstein polynomials for core mass and initial atmospheric mass fraction, as well as a Gaussian distribution for core composition is labelled as \textsc{model III}.

\subsection{Generating the Planet Sample} \label{sec:DrawingPlanets}
\begin{figure} 
	\includegraphics[width=\columnwidth]{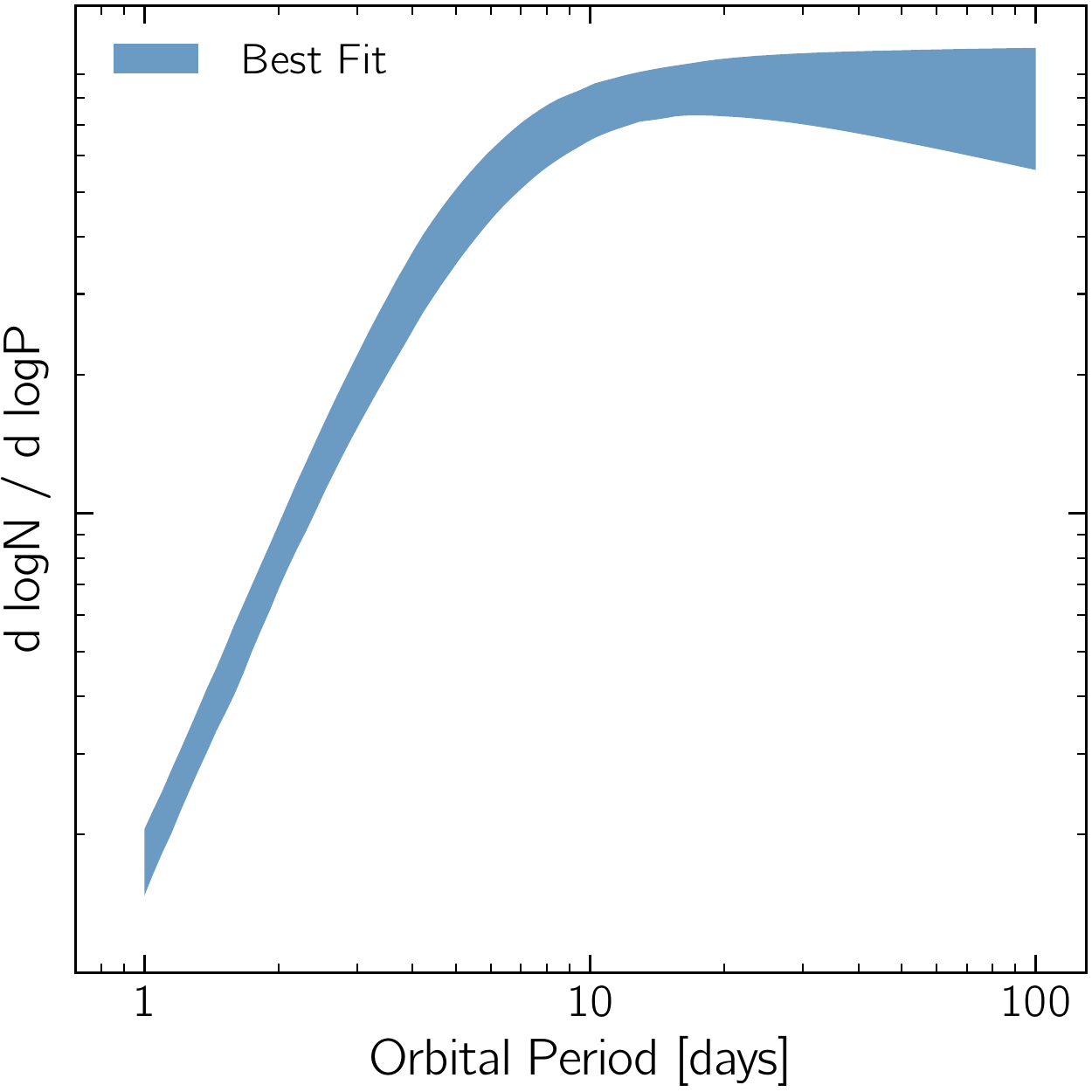}
    \caption{Best fit orbital period distribution when fit to the CKS planets, parameterised by a smooth broken power law (Equation \ref{eq:period}). Shaded region shows $1\sigma$ uncertainties. More information on this fitting can be found in Appendix \ref{sec:PeriodAppendix}.}
    \label{fig:Period} 
\end{figure}

In addition to core mass, initial atmospheric mass fraction and composition, we require values for host stellar mass $M_*$ and orbital period $P_\text{orb}$ in order to model a planet's atmospheric evolution. Whilst, strictly speaking, these distributions should be inferred as part of our full hierarchical model, we choose not to include them for a variety of reasons. Firstly, stellar mass measurements are essentially decoupled as they are directly measured as part of the CKS program. One can therefore fit this distribution independently without affecting the other distributions. The orbital period distribution on the other hand is coupled, albeit weakly, with the other model parameters, meaning it is more important to include in the full inference problem. However, when this was attempted, the MCMC chain became stuck in local minima, resulting in uninformative posteriors and computationally infeasible runtimes. As this removed the possibility of fitting the period distribution in the full model, we choose to fit it independently, drawing upon multiple previous works that achieve the same task \citep[e.g.][]{Fressin2013,Howard2012,Petigura2018}. The added benefit of not including the fit of either stellar mass of orbital period distributions is that is it reduces the number of model parameters and hence simplifies the inference problem. As previously stated, we choose to infer exoplanet demographic properties by comparing synthetic measurements with real observations. To this end, we fit the stellar mass and orbital period distribution by forward modelling the detections of such quantities, given an underlying population. For the orbital period, motivated by previous works \citep[e.g.][]{Fressin2013,Howard2012,Dressing2015b,Petigura2018}, we choose our underlying distribution to be a smooth broken power law:
\begin{equation} \label{eq:period}
    \frac{\mathrm{d}N}{\mathrm{d}P} \propto \frac{1}{\big(\frac{P}{P_0}\big)^{-k_1} + \big(\frac{P}{P_0}\big)^{-k_2}},
\end{equation}
where we find the power law break at $P_0 = 5.75$ days, with $k_1 = 2.31$ and $k_2 = -0.08$, as shown in Figure \ref{fig:Period}. Whilst completeness effects are included in this fitting, we choose not to add noise to the modelled orbital periods as the fractional error from \textit{Kepler} is typically $\sim 10^{-6}$ and hence negligible. More details of this fit can be found in Appendix \ref{sec:PeriodAppendix}.

When fitting the host stellar mass distribution we choose an underlying Gaussian function and add noise to the data in the form of Gaussian perturbations that are typical of the CKS catalogue. The best fit and hence the distribution we draw our host stellar masses from is:
\begin{equation}
    \frac{\mathrm{d}N}{\mathrm{d} M_*} \propto \exp{ \frac{-(M_* - \mu_{M_*})^2}{2 \sigma^2_{M_*}}}.
\end{equation}
where we find $\mu_{M_*}=1.04M_\odot$ and $\sigma_{M_*}=0.15M_\odot$. This distribution is shown in Figure \ref{fig:Smass}, with further information found in Appendix \ref{sec:MstarFitting}. Including these fits into our model allows us to evolve each planet with a core mass, initial atmospheric mass fraction, core density, host stellar mass and orbital period drawn from their respective distributions.

\begin{figure} 
	\includegraphics[width=\columnwidth]{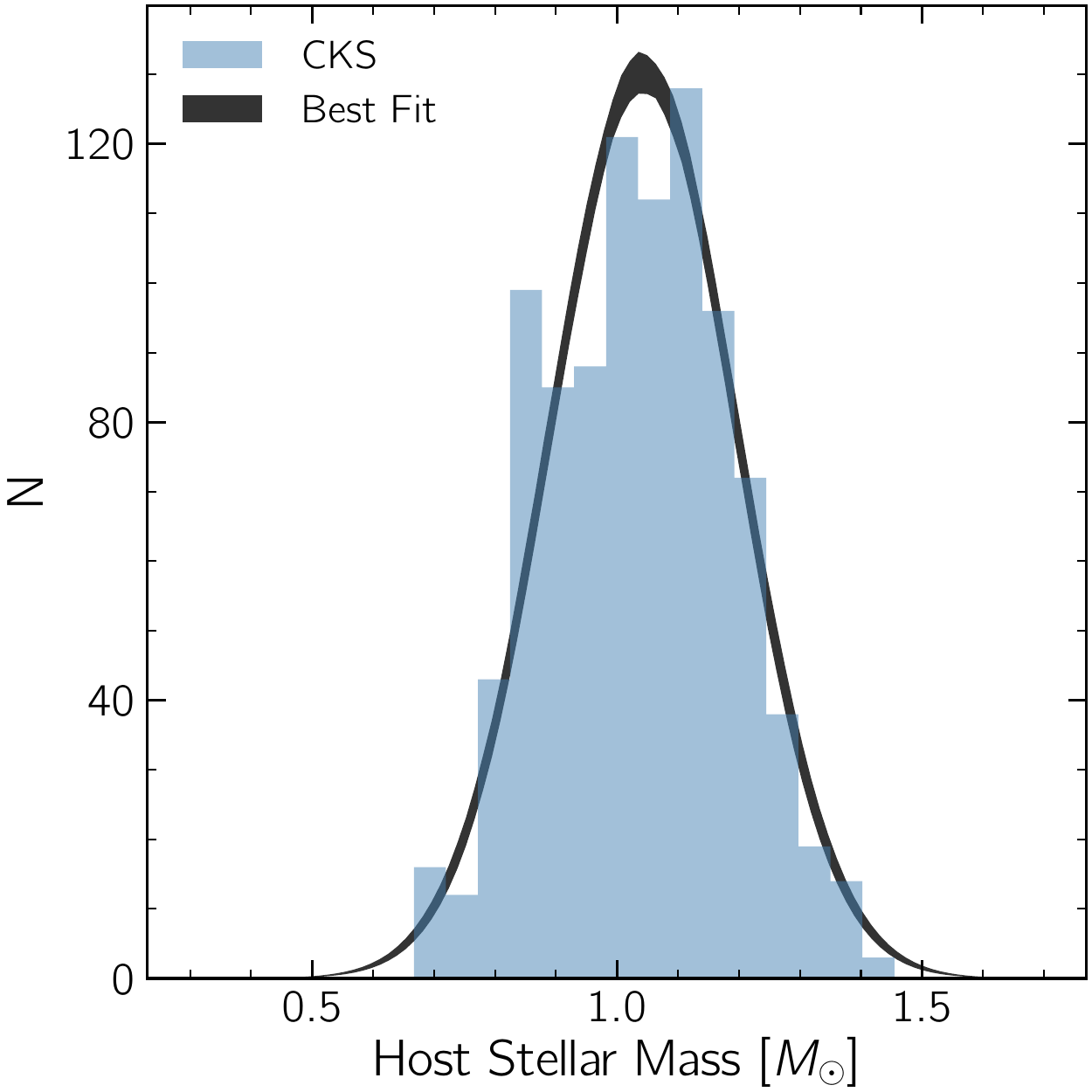}
    \caption{A histogram of CKS host stellar masses is plotted with best fit Gaussian distribution on top. This has mean $\mu_{M_*}=1.04M_\odot$ and standard deviation $\sigma_{M_*}=0.15M_\odot$. The width of the best-fit line represents $1\sigma$ uncertainties. More detail can be found in Appendix \ref{sec:SmassAppendix}.}
    \label{fig:Smass} 
\end{figure}

\subsection{Evolving the Planets} \label{sec:EvolvingPlanets}
\begin{figure} 
	\includegraphics[width=\columnwidth]{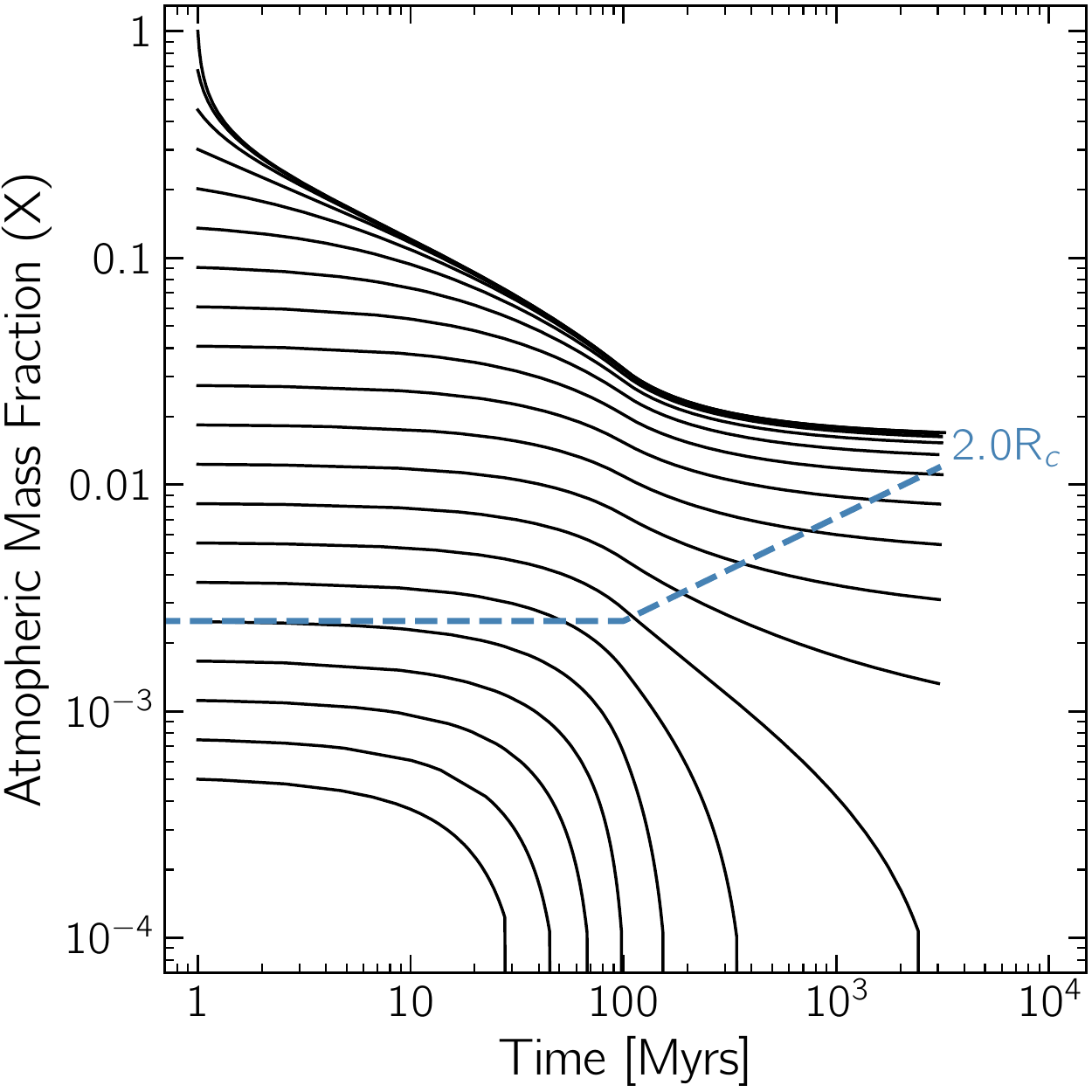}
    \caption{Solid lines show the evolution of varying initial atmospheric mass fractions for a $5M_\oplus$ core with Earth-like composition, orbiting a Sun like star at $10$ day orbital period. Calculations are performed using the analytic model of \protect\cite{OwenEstrada2020}. The dashed line tracks an atmospheric mass fraction required to double the core radius. This plot reproduces results from \protect\cite{Owen2017} in which a planet either loses its gaseous atmosphere within $100$ Myr, or retains an atmospheric mass fraction $\sim 1\%$ which approximately doubles its core radius. This bimodal evolution essentially forms the basis for the origin of the radius gap \protect\citep[see][]{Owen2019}.}
    \label{fig:XvsT}
\end{figure}

The evolution of the planets' atmosphere arises in this model due to EUV/X-ray photoevaporation and cooling. In \cite{Owen2017}, a planet's photospheric radius is calculated by approximating the evolution of the H/He atmosphere, allowing one to evaluate the photospheric radius of a exoplanet as a function of time. In this work however, we adopt the methodology of \cite{OwenEstrada2020}, that relaxed some of the assumptions of \citet{Owen2017} in determining the planet's radius. We do note that our planetary structure and evolutionary calculations are still approximate, and we are not solving the full stellar structure equations \citep[unlike, e.g.][]{Owen2013,OwenMorton2016,ChenRogers2016}; however, in this work we perform approximately $\sim 10^{10}$ planetary evolution calculations, something only possible with the simplified scheme. Within this formalism,  calculating the atmospheric evolution involves solving an ordinary differential equation for the evolution of a planet's atmospheric mass fraction $X \equiv M_{atm} / M_{core}$:
\begin{equation} \label{eq:ODE}
    \frac{\mathrm{d}X}{\mathrm{d}t} = -\frac{X}{t_{\dot{X}}},
\end{equation}
where $t_{\dot{X}}$ is the atmospheric mass-loss timescale, given by:
\begin{equation}
    t_{\dot{X}} \equiv \frac{X}{\dot{X}} = \frac{M_{atm}}{\dot{M}_{atm}}.
\end{equation}
The mass-loss rate $\dot{M}_{atm}$ is calculated using the energy-limited mass-loss model \citep[e.g.][]{Baraffe2004,Erkaev2007}, which provides a self-consistent method for calculating the photospheric radius $R_{p}$ of a planet given it's atmospheric mass fraction:
\begin{equation} \label{eq:massloss}
    \dot{M}_{atm} = \eta \; \frac{\pi R_\text{p}^3 L_\text{XUV}}{4 \pi a^2 GM_\text{p}},
\end{equation}
where $a$ is the orbital semi-major axis, $L_\text{XUV}$ is the high energy luminosity from the host star and $\eta$ is the mass-loss efficiency. To quantify this efficiency, we adopt an approximate fit to mass-loss simulations from \cite{OwenJackson2012}, as used in \cite{Owen2017,Wu2019}:
\begin{equation} \label{eq:masslossefficiency}
    \eta = \eta_0 \; \left( \frac{v_\text{esc}}{25\text{km s}^{-1}} \right ) ^{-\alpha_\eta}
\end{equation}
where $v_\text{esc}$ is the escape velocity of the planet, $\eta_0$ is the normalisation and $\alpha_\eta$ is the power-law index. Whilst the value for $\eta_0$ is taken to be $0.17$; motivated by hydrodynamic simulations \citep{OwenJackson2012}, we choose to add $\alpha_\eta$ to our inference parameters in \textsc{model III}. The reason for not incorporating $\eta_0$ is due to a strong degeneracy shared with core composition \citep{OwenAdams2019}, which would require either perfect numerical photoevaporation models, or planet mass data (either RV or TTV) in order to break. Therefore, as described in \citet{OwenAdams2019,Mordasini2020}, our constraints on core-composition are completely dependent on the accuracy of photoevaporation simulations; as discussed further in Section \ref{sec:Discussion}. In practise, Equation \ref{eq:ODE} is solved using an RK45 adaptive-step numerical integrator with a error tolerance of $10^{-5}$, which can handle the sharp changes in $X$ as seen in Figure \ref{fig:XvsT}. Note that we assume a planet has been completely stripped when it's atmospheric mass fraction falls below $10^{-4}$, a value at which the planet's radius is indistinguishable from the core's radius in transit observations.

As shown in \cite{Wright2011,Jackson2012,Tu2015}, stellar $L_\text{XUV} / L_\text{bol}$ decays quickly once a star begins to spin down and the production of high-energy photons weakens, which typically happens at $\sim 100\text{ Myr}$ for sun-like stars. In our model, we take the $L_\text{XUV}$ magnitude and evolution from \cite{Owen2017}:
\begin{equation}
    L_\text{XUV}=\begin{cases}
    L_\text{sat} & \text{for } t < t_\text{sat}, \\
    L_\text{sat} \bigg ( \frac{t}{t_\text{sat}} \bigg)^{-1-a_0}  & \text{for } t \geq t_\text{sat},
    \end{cases}
\end{equation}
where $a_0=0.5$, $t_\text{sat}=100 \text{ Myr}$ and the saturation luminosity follows
\begin{equation}
    L_\text{sat} \approx 10^{-3.5} L_\odot \bigg( \frac{M_*}{M_\odot} \bigg),
\end{equation}
which is motivated from both observational and theoretical studies \citep[e.g.][]{Gudel1997,Ribas2005,Wright2011,Jackson2012,Tu2015}. Crucially, the time-scale for this decay is far shorter than atmospheric mass-loss timescales for small, close-in exoplanets. In our model, each planet is evolved for $3$Gyr to match the typical ages of Kepler systems. Recall that the majority of mass-loss occurs during the first few $100$ Myr, so changing the total evolution time by even Gyrs has little to no effect on the final size distribution of the simulated planets as seen in Figure \ref{fig:XvsT}. 

\begin{figure*} 
	\includegraphics[width=2.1\columnwidth]{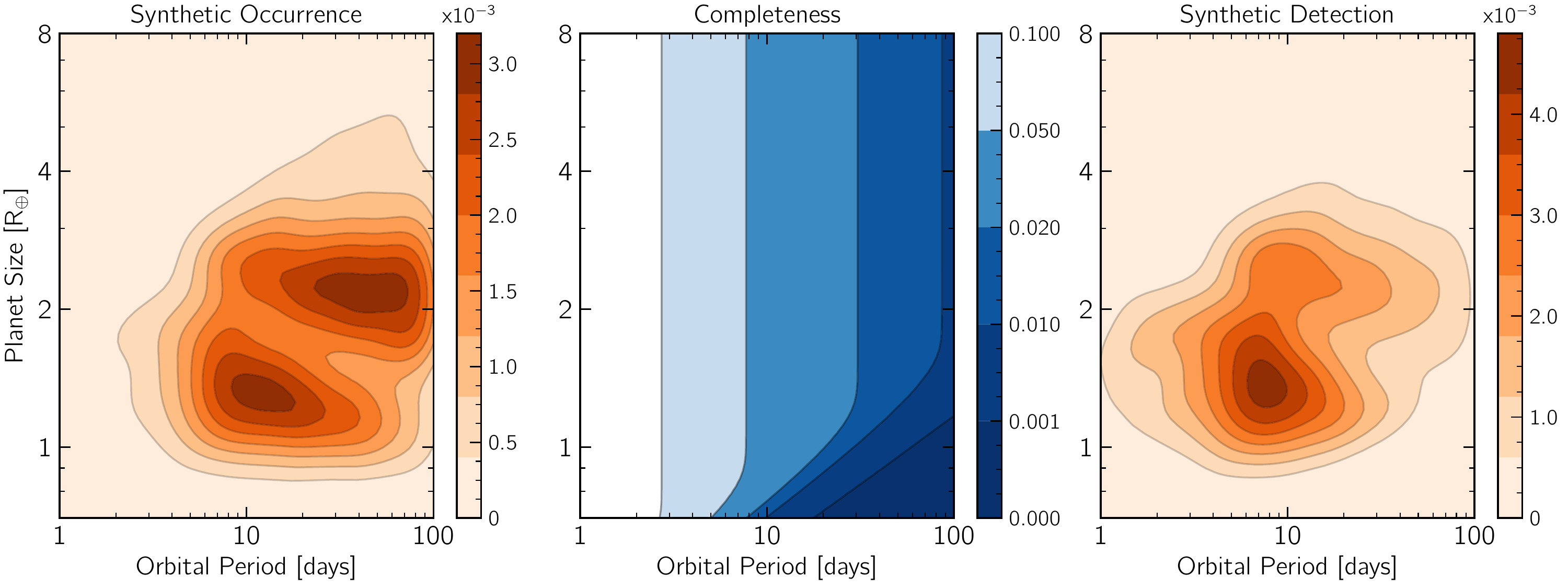}
    \caption{The left-hand panel shows an example PDF for synthetic occurrence of exoplanets in the orbital period - radius plane. This is produced by forward-modelling an ensemble of planets through photoevaporation, producing the bimodal distribution of super-Earths and sub-Neptunes. The central panel shows the completeness of the Kepler survey in the same domain. Contours represents the product of $p_\text{transit} \cdot p_\text{det}$, which is increased for close-in, large planets and reduced for far-out, small planets. In the right-hand panel we show the synthetic detection PDF $\lambda(P,R_\text{p})$, which is the product of the the occurrence and the completeness and represents the planet occurrence weighted by the Kepler survey biases. One can see that planets with a large orbital period and hence a small $p_\text{transit}$ are heavily disfavoured to be detected. }
    \label{fig:completeness}
\end{figure*}

\subsection{Detecting the Planets} \label{Sec:Detection}
Once the planets have been evolved through 3 Gyr of photoevaporation, we perform a synthetic \textit{Kepler} transit survey with the intention of then comparing it with real CKS data. We add errors to the simulated photospheric radii of the evolved planets as well as introduce observational biases that match the real data. To quantify the radii errors of the real data, we take the typical fractional error of the CKS planets ($\sim 6\%$) and add a similar fractional error to our evolved planet radii by adding a random Gaussian perturbation with zero-mean and standard deviation of $\sigma=0.06$. This has the effect of adding noise to the data - which shifts some planets into the radius valley, as opposed to a clean valley as seen in asteroseismic surveys \citep{VanEylen2018}, see \citet{Petigura2020}. We choose not to add fractional errors to the orbital period measurements as these were measured with extreme precision by \textit{Kepler}, typically $\sim 10^{-6}$. As a result, the main source of noise in the CKS data set is from the planet size measurements, as well as sampling uncertainty (Poisson shot noise).

{\correction We must now introduce the biases associated with the \textit{Kepler} survey. The most rigorous method to achieve this would be to directly inject the modelled planets into the \textit{Kepler} pipeline and determine whether they would have been detected around the CKS stars. However, this task is computationally unfeasible for a large hierarchical model such as this. We therefore decide to take an average approach,} following the prescription from \cite{Fulton2017}, in which forward models of planets are biased by their associated probability of being detected. The first contribution to this derives from the geometric probability of transit:
\begin{equation}
    p_\text{transit} = b_\text{cut} \; \frac{R_*}{a},
\end{equation}
where $b_\text{cut}=0.7$ represents the cutoff in transit impact parameter chosen in \cite{Fulton2017} to avoid grazing transits. The second contribution to the completeness is the probability of detection, resulting from the Kepler pipeline efficiency:
\begin{equation}
    p_\text{det} = \frac{1}{N_*} \sum^{N_*}_i C(m_i)
\end{equation}
where $N_* = 36,075$ is the number of stars in the Stellar17 catalogue\footnote{\url{https://archive.stsci.edu/kepler/stellar17/search.php}} \citep{Mathur2017} that pass the \textit{Kepler} pipeline \citep{Christiansen2012,Christiansen2015,Christiansen2016} as well as CKS filters. The fraction of injections $C$ recovered in the \textit{Kepler} pipeline is a function of injected signal-to-noise $m_i$. As in \cite{Fulton2017}, we choose the pipeline efficiency $C(m_i)$ to be a $\Gamma$ cumulative distribution function of form,
\begin{equation}
    C(m_i; k,\theta,l) = \Gamma(k) \int^{\frac{m_i-l}{\theta}}_{0} t^{k-1} \; e^{-t} \; \mathrm{d} t,
\end{equation}
with $k=17.56$, $l=1.00$ and $\theta=-0.49$. The product of $p_\text{transit} \cdot p_\text{det}$ is used is calculated to produce a mean completeness map for the entire CKS sample (similar to Figure 6 in \cite{Fulton2017}) that captures the completeness probability of the survey for a given $P_\text{orb}$ and $R_\text{p}$. This is shown in the centre panel of Figure \ref{fig:completeness}.

In order to introduce the completeness map to the evolved and noisy (i.e. including $R_\text{p}$ errors) synthetic planet data, we adopt the following process: first we perform a 2D Gaussian Kernel Density Estimation (KDE) on the synthetic planet population in $P_\text{orb}$-$R_\text{p}$ space. Crucially, the bandwidth of the KDE generator $\sigma_\text{KDE} = 0.05$ is chosen to be less than typical planet radii uncertainty, as this ensures the behaviour of the function is set by the planet distribution and not numerical effects from the construction of the KDE (discussed in Section \ref{sec:Convergence}, see Figure \ref{fig:convergence}). This KDE can be thought to be an approximation to the PDF for planet occurrence. We then bias this PDF with the \textit{Kepler} completeness map and normalise to unity, which results in a PDF for planet detections, which we label $\lambda(P_\text{orb}, R_\text{p})$. In other words, this PDF represents the probability of detecting a planet at a given orbital period and planet radius, given a set of model parameters. An example of synthetic occurrence and synthetic detection PDFs are shown in Figure \ref{fig:completeness}. This PDF is then used as the basis for our Likelihood calculation.

\subsection{Likelihood Calculation} \label{sec:Likelihood}
The final step in the inference problem is to compare the real CKS data to the noisy, incomplete synthetic data. As used in previous works \citep[e.g.][]{ForemanMackey2014,Hsu2018,Bryson2020}, we assume that the detection of an exoplanet is an independent process. Thus, we model the detection of an exoplanet with the transit method as an inhomogeneous Poisson point process. We can calculate the total occurrence factor $\Lambda$:
\begin{equation} \label{eq:Lambda}
    \Lambda = N_* \cdot f_* \cdot \iint \lambda(P,R_\text{p}) \: \mathrm{d} P \, \mathrm{d} R_\text{p},
\end{equation}
where $N_* = 36,075$ is again the number of stars in the filtered Stellar17 catalogue and $f_*$ is the mean number of planets per star in the \textit{Kepler} field. We allow $f_*$ to vary as a model parameter in all simulations in the aim of inferring its value. Whilst in our model, $f_*$ is a nuisance parameter, other works have derived this value in far more detail \citep[e.g.][]{Howard2012,Fulton2017,Zhu2018,Zink2019,Hsu2019}. The probability of detecting $n$ planets is:
\begin{equation}
    P(n) = e^{-\Lambda} \, \frac{\Lambda^n}{n!},
\end{equation}
whereas the probability density of observing a planet at $(P_i, R_{\text{p},i})$ is:
\begin{equation} \label{eq:ProbDens}
    \rho\,(P_i, R_{\text{p},i}) = \frac{\lambda(P_i, R_{\text{p},i})}{\Lambda}.
\end{equation}
Assuming that all observations are independent\footnote{In this calculation, we are thus assuming single and multiple transiting systems are implicitly drawn from the same formation model. As a result this ignores the detection bias that occurs in multi systems \citep{Zink2019}. }, the probability of detecting an ensemble of planets at $\pi = \{(P_1, R_{\text{p},1}), ... , (P_n, R_{\text{p},n}) \}$ is:
\begin{equation}
    P(\pi) = \prod_i^n \frac{\lambda(P_i, R_{\text{p},i})}{\Lambda}.
\end{equation}
The likelihood of observing this sample is thus:
\begin{equation}
    L(\pi) = P(n) \cdot P(\pi) = e^{-\Lambda} \, \frac{\Lambda^n}{n!} \; \prod_i^n \frac{\lambda(P_i, R_{\text{p},i})}{\Lambda}.
\end{equation}
In this hierarchical inference model, we wish to sample a posterior with the assumption that the ensemble of planets $\pi$ is the CKS sample, whilst the detection rate $\lambda(P,R_\text{p})$ is calculated using a forward modelled sample of planets drawn from distributions defined by our model parameters $\boldsymbol{\theta}$. As a result, the final likelihood is:
\begin{equation} \label{eq:likelihood}
\begin{split}
L(\pi_\text{CKS}) & = P(N_\text{CKS}) \cdot P(\pi_\text{CKS}) \\
 & = e^{-\Lambda} \, \frac{\Lambda^{N_\text{CKS}}}{N_\text{CKS}!} \; \prod_i^{N_\text{CKS}} \frac{\lambda(P_i, R_{\text{p},i})}{\Lambda}.
\end{split}
\end{equation}
where $N_{\text{CKS}}$ is the number of planets in the CKS sample and $\pi_\text{CKS}$ is the set of values of $(P_\text{orb}, R_\text{p})$ for all CKS planets within our chosen domain. Similarly to \cite{Fulton2017}, we restrict the orbital period to lie between $1$ and $100$ days, whilst the planet size at 3~Gyr can vary between $0.95R_\oplus$ and $6R_\oplus$. The lower bound here is chosen to avoid areas of low-completeness which are almost devoid of planets\footnote{More specifically, this lower limit of $0.95R_\oplus$ is chosen as a compromise between inferring the core mass distribution to smaller masses whilst still allowing smaller cores to have an icy composition and still be detected.}, whilst the upper bound is chosen to avoid modelling large planets hosting self-gravitating atmospheres which are not accounted for in the evolution model and may form in a different way \citep[e.g.][]{OwenLai2018}. Our likelihood function (Equation \ref{eq:likelihood}) can be interpreted in two parts: the Poisson pre-factor is maximised when the number of synthetically observed planets from the model matches the number observed in the CKS data set. Therefore this factor allows us to put realistic uncertainties on our distribution of interest due to the finite sampling of the exoplanet data. The product factor on the other hand is maximised when the shape of the modelled planet distribution in $P_\text{orb}$-$R_\text{p}$ space matches that of the CKS planets. We therefore find a high likelihood when the number of observed planets and shape of the planet distribution match the CKS data.

In order to sample from our posterior, the log-likelihood (i.e. $\ln L(\pi_\text{CKS})$) is fed into a Monte Carlo Markov chain (MCMC) algorithm, specifically the \textsc{emcee} Python implementation \citep{emcee} of the affine-invariant ensemble sampler from \cite{GoodmanWeare2010}. We run with 150 walkers and a sufficient number of iterations required for chain convergence, which varies for each of the models and is discussed further in the supplementary material.

\subsubsection{Likelihood Convergence} \label{sec:Convergence}
\begin{figure} 
	\includegraphics[width=\columnwidth]{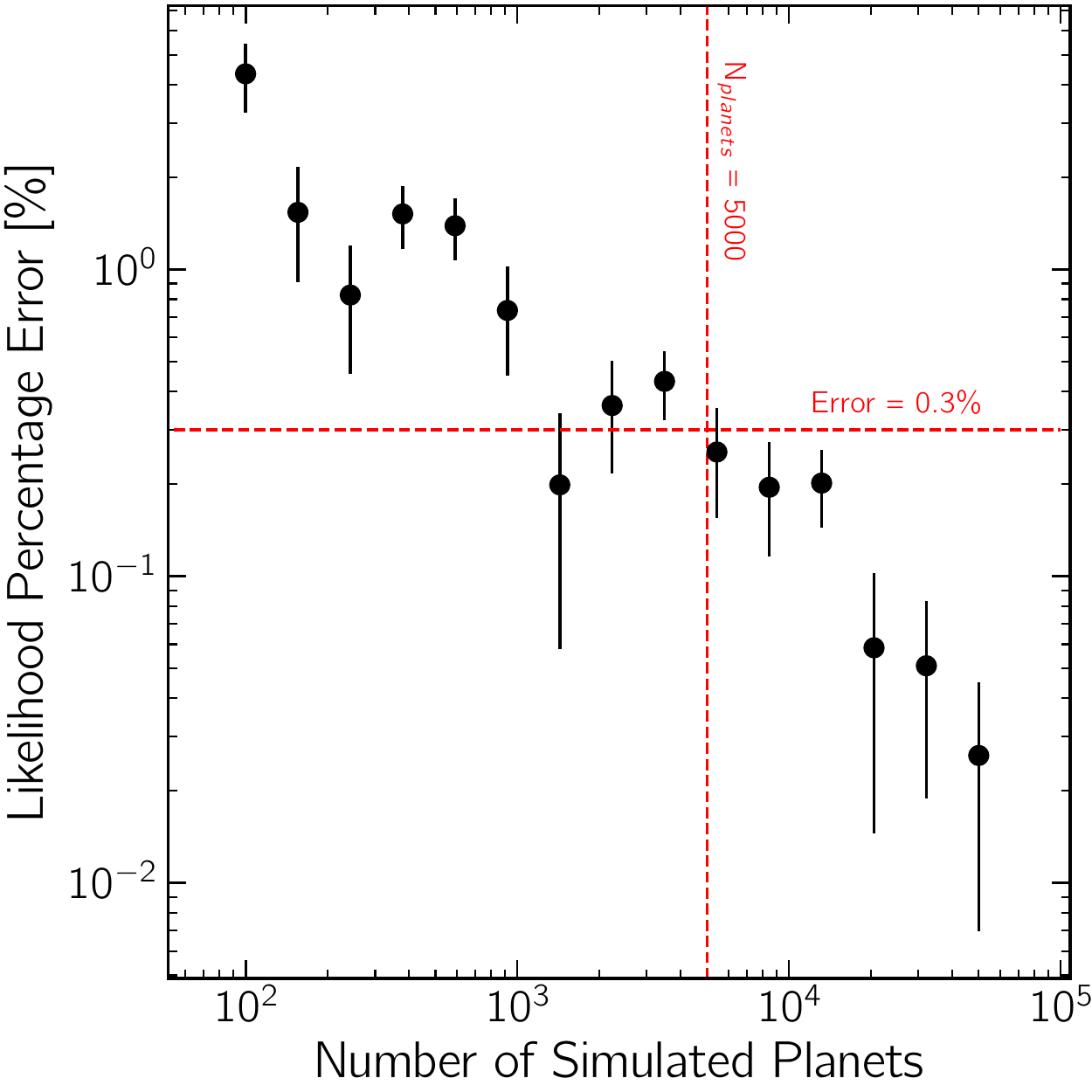}
    \caption{Percentage error in log-likelihood is plotted as a function of number of planets simulated for \textsc{model II}. The stochastic nature of the model introduces noise into error calculations, which is quantified by repeating each simulation 100 times and determining the standard deviation for each value of simulated planets. Dotted lines represents our adopted compromise between likelihood accuracy and computational speed, with a percentage error of $0.3\%$, achieved by simulating 5000 planets per MCMC iteration.}
    \label{fig:convergence} 
\end{figure}

In order to ensure our planet occurrence PDF $\lambda(P,R_\text{p})$, and hence likelihood is accurate and converged, we require a suitably large number of planets to be simulated for each iteration of the MCMC. This is so that the planet radii errors we add in the synthetic detection process are integrated over and the PDF for planet occurrence/detections are controlled by the planet distribution shape, and not subject to sampling noise. On the other hand however, simulating a large number of planets (albeit semi-analytically) can become a computationally expensive process. We therefore wish to find a compromise. Figure \ref{fig:convergence} shows convergence tests for \textsc{model II}. This is done by calculating the percentage difference of the likelihood from the `true' value for a range of simulated planets. We approximate the true likelihood by evaluating our likelihood (Equation \ref{eq:likelihood}) with $100,000$ planets and thus achieving an extremely high accuracy.

The error in the log-likelihood changes as a power-law with increasing number of planets indicating convergence. We choose to use 5000 planets as this provides an error of $\lesssim 0.3 \%$ and provides a balance between accuracy and computational efficiency. 

\section{Results} \label{sec:Results}

\subsection{Model I} \label{sec:ModelI}
\textsc{model I} is the simplest of our inference problems. We choose to adopt the same parametric forms for core mass, initial atmospheric mass fraction and core composition as in \cite{Wu2019}, with our new and improved inference methodology.  We place flat priors on all parameters, with the exception of the log-Gaussian mean-values, which have log-flat priors. Values compared with that of \cite{Wu2019} are shown in Table \ref{tab:modelI}.
\begin{table}
	\centering
	\caption{Comparing constrained parameters from \textsc{model I} and \protect\cite{Wu2019}, in which core mass and initial atmospheric mass fraction are restricted to be log-Gaussian. Here, $\mu_\text{M}$ and $\sigma_\text{M}$ are mean and standard deviation for core mass distribution, whilst $\mu_\text{X}$ and $\sigma_\text{X}$ are mean and standard deviation for initial atmospheric mass fraction distribution (Equation \ref{eq:logGauss}). Quoted errors for \textsc{model I} are the 1$\sigma$ percentiles calculated from marginalised posteriors.}
	\def\arraystretch{1.5}
	\begin{tabular}{ lcr } 
		\hline
		Parameter & \textsc{model I} & \cite{Wu2019}\\
		\hline
		$\mu_\text{M} \; [M_\oplus]$ & $3.72^{+0.45}_{-0.33}$ & $7.70^{+1.50}_{-1.50}$\\
		$\sigma_\text{M}$ & $0.44^{+0.06}_{-0.03}$ & $0.29^{+0.06}_{-0.06}$\\
		$\mu_X$ & $0.040^{+0.015}_{-0.016}$ & $0.026^{+0.006}_{-0.006}$\\
		$\sigma_X$ & $0.51^{+0.20}_{-0.12}$ & $0.29^{+0.06}_{-0.06}$\\
		\hline
	\end{tabular}
	\label{tab:modelI}
\end{table}
We note that whilst our adopted methodologies are similar, there are distinct differences, hence we expect to see difference in the results. We therefore include this comparison to demonstrate how different results are found with different methodologies. One of the main differences which likely has the largest effect is that our model infers a 2D planet distribution, as opposed to a 1D radius histogram, which likely widens our constrained distributions. In addition, we adopt a more accurate completeness model and hence allow a larger fraction of smaller planets to be detected - which results in a lower mean core mass. Finally, in \cite{Wu2019}, core masses were scaled with stellar mass which will also act to narrow constrained distributions (further discussed in Section \ref{sec:Discussion}). This comparison indicates that performing this inference in the 2D planet distribution is important.

\subsection{Model II} \label{sec:ModelII}

\textsc{model II} relaxes the condition of a specific functional form for the core masses and initial atmospheric mass fractions. For both distributions, we use a $5^\text{th}$-order Bernstein polynomial expansion with model parameters being the $5$ polynomial coefficients. As with \textsc{model I}, we also constrain a single value for core composition, and the mean number of planets per star. The left hand panel of Figure \ref{fig:Combined-models} demonstrates a good agreement of model and data by comparing radius histograms. This is produced by resampling 1000 sets of model parameters from the MCMC chain and evolving 5000 planets for each sample. By then calculating a PDF for planet detection (as laid out in Section \ref{Sec:Detection}), we can determine the observed radius distribution and thus $1\sigma$ errors across the resampled models. \textsc{model II} can be seen to reproduce the bimodal distribution with peaks and gap in the correct positions. 

The best fit core mass and initial atmospheric mass fraction distributions, produced from their associated Bernstein polynomial are shown in the middle and right-hand panel of Figure \ref{fig:Combined-models}. Similar to \cite{Wu2019}, we find a peaked core mass distribution with most planets hosting a core of a few earth masses. Crucially, we see this distribution is not symmetric, unlike a log-Gaussian, with a positive skewness of $0.04\pm0.01$ and kurtosis of $0.15^{+0.02}_{-0.03}$. This result thus confirms the requirement of relaxing assumptions from \textsc{model I}. Similarly, and starkly different to the latter, the initial atmospheric mass fraction distribution is asymmetric and strongly peaked at $\sim 10^{-2}$, implying that the majority of Kepler planets were born with a significant H/He atmosphere of $\sim 1\%$ mass, instead of being born with a negligible atmosphere. The skewness and kurtosis of this distribution are $-0.16^{+0.12}_{-0.09}$ and $1.45^{+0.21}_{-0.13}$ respectively.

In addition to core mass and initial atmospheric mass fraction, we also constrain the mean number of planets per star to be $f_* = 0.72\pm0.03$ and the core composition to be $\tilde{\rho} = 0.18\pm0.09$. This composition equates to a bulk density for a $1M_\oplus$ core of $\rho_{M_\oplus} = 4.74^{+0.40}_{-0.36} \text{ g cm}^{-3}$, pertaining to a slightly lower density to that of Earth.

\subsection{Model III} \label{sec:ModelIII}

\begin{figure} 
	\includegraphics[width=\columnwidth]{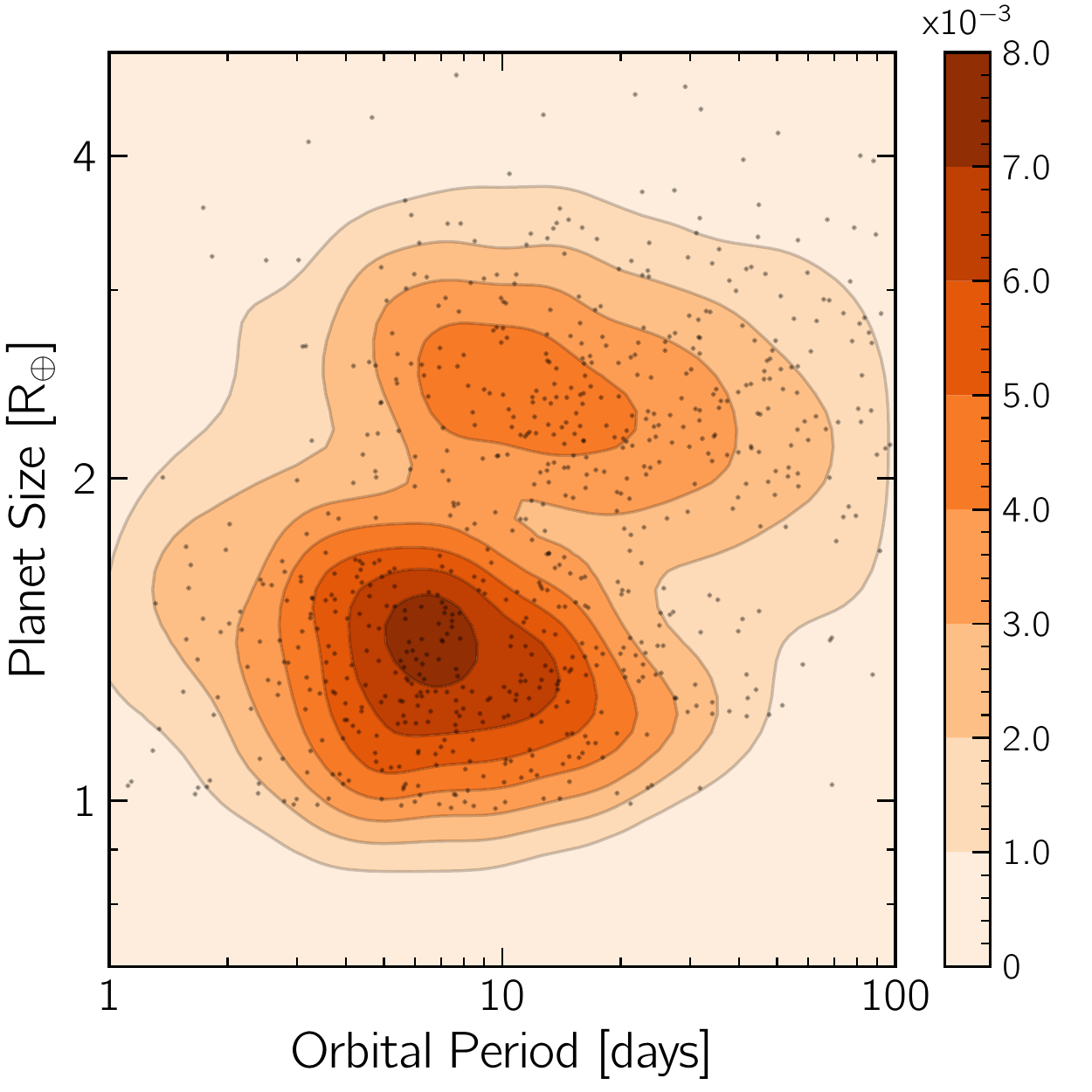}
    \caption{Synthetic detection PDF is shown in orange contours for best fit parameters from \protect\textsc{model III}. CKS planets from \protect\cite{Fulton2017} are also shown to demonstrate the goodness of fit of model and data.}
    \label{fig:detection_MIII} 
\end{figure}

\begin{figure*} 
	\includegraphics[width=2.1\columnwidth]{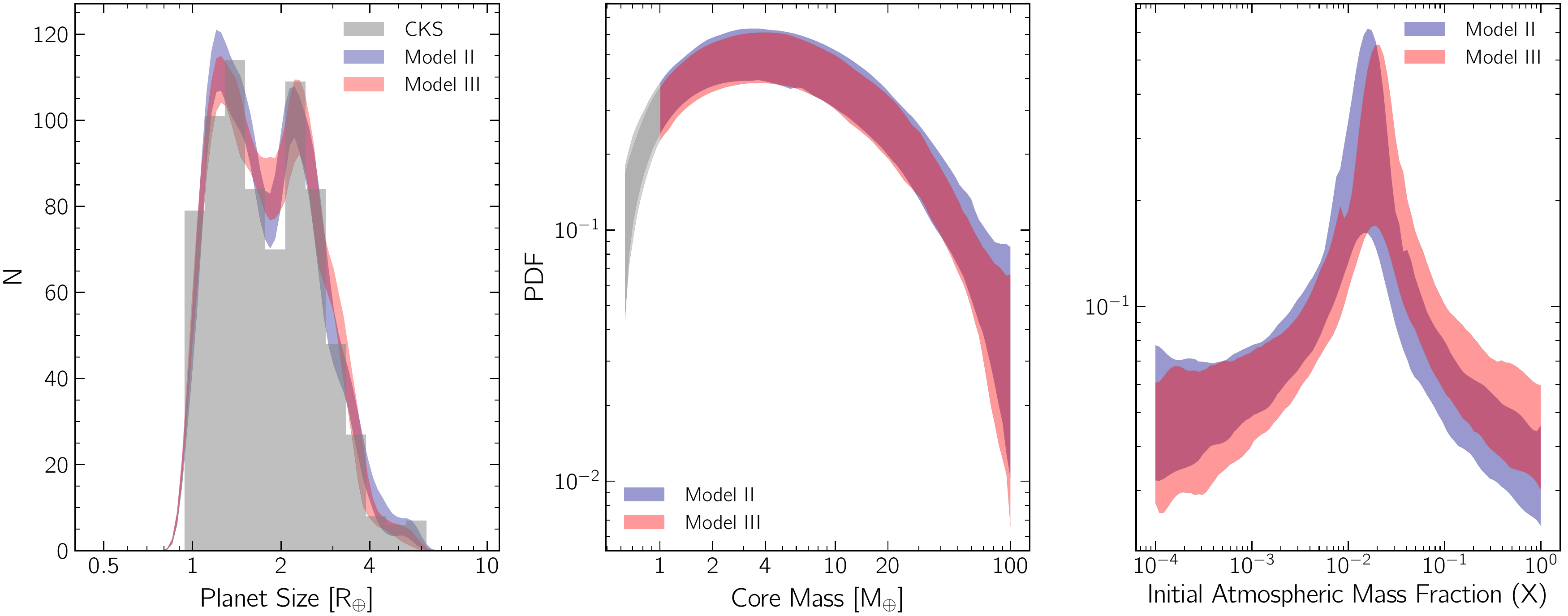}
    \caption{Left hand panel shows radius histogram comparison of CKS data with best fit from \protect\textsc{model II} and \protect\textsc{model III}. The shaded line represents $1\sigma$ errors of $1000$ forward-models resampled from the MCMC chains, whilst grey histogram bins are the CKS data from \protect\cite{Fulton2017}. Middle and right panels show constrained core mass and initial atmospheric mass fraction distribution for \protect\textsc{model II} and \protect\textsc{model III}, both of which are described by $5^\text{th}$-order Bernstein polynomial expansions. Shaded regions are representative of a $1\sigma$ confidence interval, calculated from 1000 resamples from the MCMC chains. Grey region in core mass distribution represents masses $<1M_\oplus$ and hence under the current detection limit.}
    \label{fig:Combined-models} 
\end{figure*}

\begin{figure*} 
	\includegraphics[width=2.1\columnwidth]{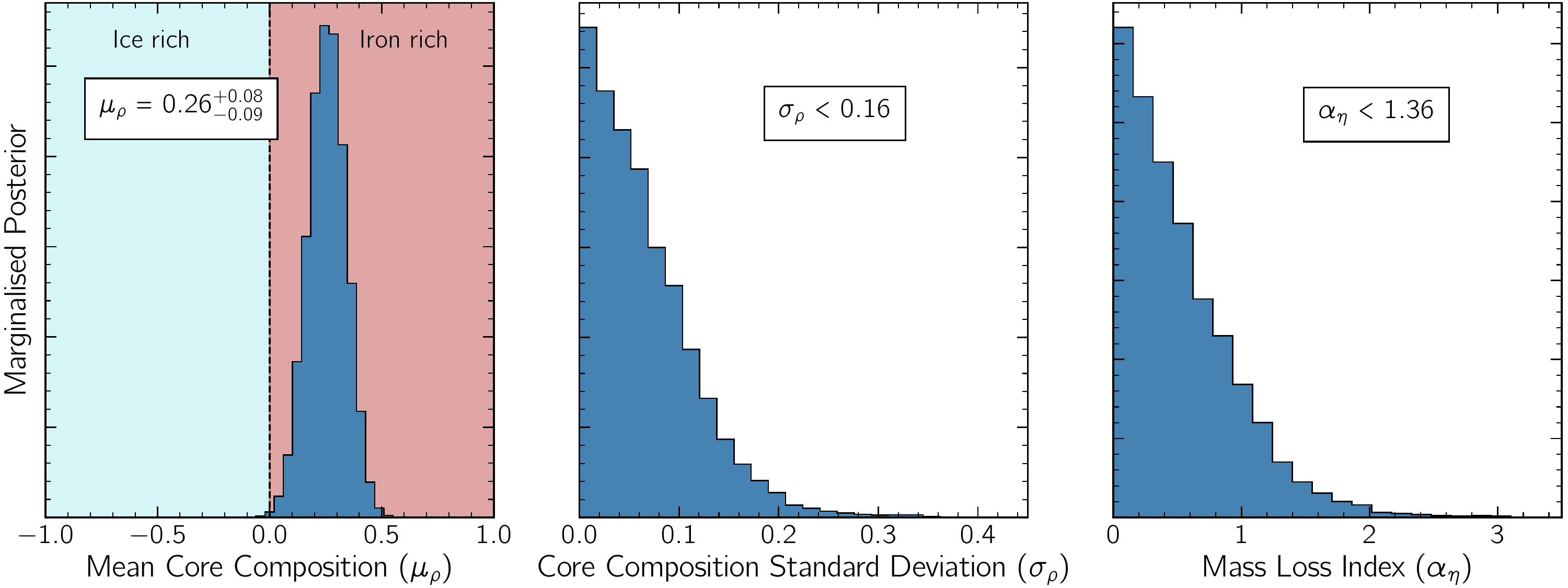}
    \caption{Marginalised posteriors from \textsc{model III}. Left-hand panel shows the mean $\mu_\rho$ for the Gaussian function used to fit the core composition distribution, with $1\sigma$ uncertainties quoted. The area $\mu_\rho<0$ represents the region of parameter space corresponding to ice rich cores, whilst $\mu_\rho>0$ corresponds to iron rich cores. Middle panel shows posterior for Gaussian composition distribution standard deviation, whilst right-hand panel shows posterior for photoevaporation mass-loss index from Equation \ref{eq:masslossefficiency}. Note that due to uniform priors on core composition and mass-loss index, we are only able to place $2\sigma$ upper-bounds on quantities of interest.}
    \label{fig:modelIII-posteriors} 
\end{figure*}

\begin{table*}
	\centering
	\def\arraystretch{1.5}
	\begin{tabular}{ lccr } 
		\hline
		Parameter & \textsc{model I} & \textsc{model II} & \textsc{model III}\\
		\hline\hline
		Mean Number of Planets per Star ($f_*$)  & $0.70^{+0.04}_{-0.03}$ & $0.72^{+0.03}_{-0.03}$ & $0.70^{+0.03}_{-0.03}$\\
		Composition Mean & $0.27^{+0.08}_{-0.08}$ & $0.18^{+0.09}_{-0.09}$ & $0.26^{+0.08}_{-0.09}$\\
		Composition Standard Deviation & - & - & $<0.16$\\
		\hline
	\end{tabular}
	\caption{Comparison of constrained parameters from \textsc{models I, II} and \textsc{III} for core composition and mean number of planets per star in the domain of $1\leq P \leq 100 \text{ days}$ and $0.95 \leq R_p \leq 6 \; R_\oplus$. Note that \textsc{model III} is the only model which attempts to fit composition with a Gaussian function and hence constrains a value for composition distribution standard deviation.}
	\label{tab:models}
\end{table*}

\textsc{model III} is the most ambitious of our inference problems. We retain the fitting of core mass and initial atmospheric mass fraction distributions with 5$^\text{th}$ order Bernstein polynomials, but additionally choose to infer the index of photoevaporation efficiency scaling $\alpha_\eta$ (Equation \ref{eq:masslossefficiency}) assuming a uniform prior, as well as the core composition distribution with a Gaussian function (also assuming uniform prior on mean and standard deviation). We show the detection PDF $\lambda(P,R)$ for best fit parameters alongside the real CKS planets in Figure \ref{fig:detection_MIII}, which demonstrates that our adopted likelihood is effective in capturing the shape of the planet distribution. As before, the best fit radius distribution for \textsc{model III} is shown in the left hand panel of Figure \ref{fig:Combined-models}. Similar to \textsc{model II}, we get excellent agreement of model and data. As well as a mean number of planets per star of $f_* = 0.70\pm0.03$, the Bernstein coefficients (and hence distribution shape) for core mass and initial atmospheric mass fraction are consistent to $1\sigma$ between \textsc{model II} and \textsc{model III}. In particular, we find the skewness and kurtosis of the core mass distribution to be $0.04^{+0.02}_{-0.01}$ and $0.14\pm0.02$ respectively, and $-0.23^{+0.13}_{-0.08}$ and $1.52^{+0.30}_{-0.18}$ for initial atmospheric mass fraction, consistent within 1$\sigma$ of those derived in \textsc{model II}.

Figure \ref{fig:modelIII-posteriors} shows marginalised posteriors for three model parameters of interest. The left and middle panel show constraints we place on the mean and standard deviation of the composition distribution, under the assumption that it follows a Gaussian function. Consistent with \textsc{model I} and \textsc{II}, we find a mean composition of $\tilde{\rho} = 0.26^{+0.08}_{-0.09}$, which is iron-rich/water-poor and consistent with that of Earth (which has a value $\sim 0.33$). As predicted in Section \ref{sec:Distributions} and shown in Table \ref{tab:models}, we also find an extremely narrow standard deviation of the core composition distribution with $2\sigma$ upper-limit of $\sigma_{\tilde{\rho}} < 0.16$, which is evidently consistent with zero. This points towards a singular formation pathway for small, close-in exoplanets, which is discussed in Section \ref{sec:Discussion}. Finally, as shown in the right-hand panel of Figure \ref{fig:modelIII-posteriors}, we place a $2\sigma$ upper limit on the index for the photoevaporative mass-loss efficiency scaling $\alpha_\eta < 1.36$, which is consistent with the findings of \cite{Wu2019}.

\section{Discussion}\label{sec:Discussion}

Our analysis provides an illuminating view of the exoplanet population at birth. We have used the photoevaporation model to undo the billions of years of evolution experienced by the observed exoplanet population. While photoevaporation has evidently sculpted the exoplanet population and hidden its initial conditions, they have not been destroyed. In fact we have been able to leverage the photoevaporation model to uncover the ``birth'' properties of the close-in, low-mass exoplanet population. According to our results as shown in Figures \ref{fig:Combined-models} and \ref{fig:modelIII-posteriors}, the population of planets typically have cores $\sim 4M_\oplus$ and began with an atmospheric mass fraction of $\sim 1\%$. The core compositions are decidedly terrestrial, with a ratio of silicates to iron consistent with Earth's, and inconsistent with a significant amount of water. We now discuss the implication of such findings.

\subsection{Atmospheric Mass Fraction Distribution}

\begin{figure} 
	\includegraphics[width=\columnwidth]{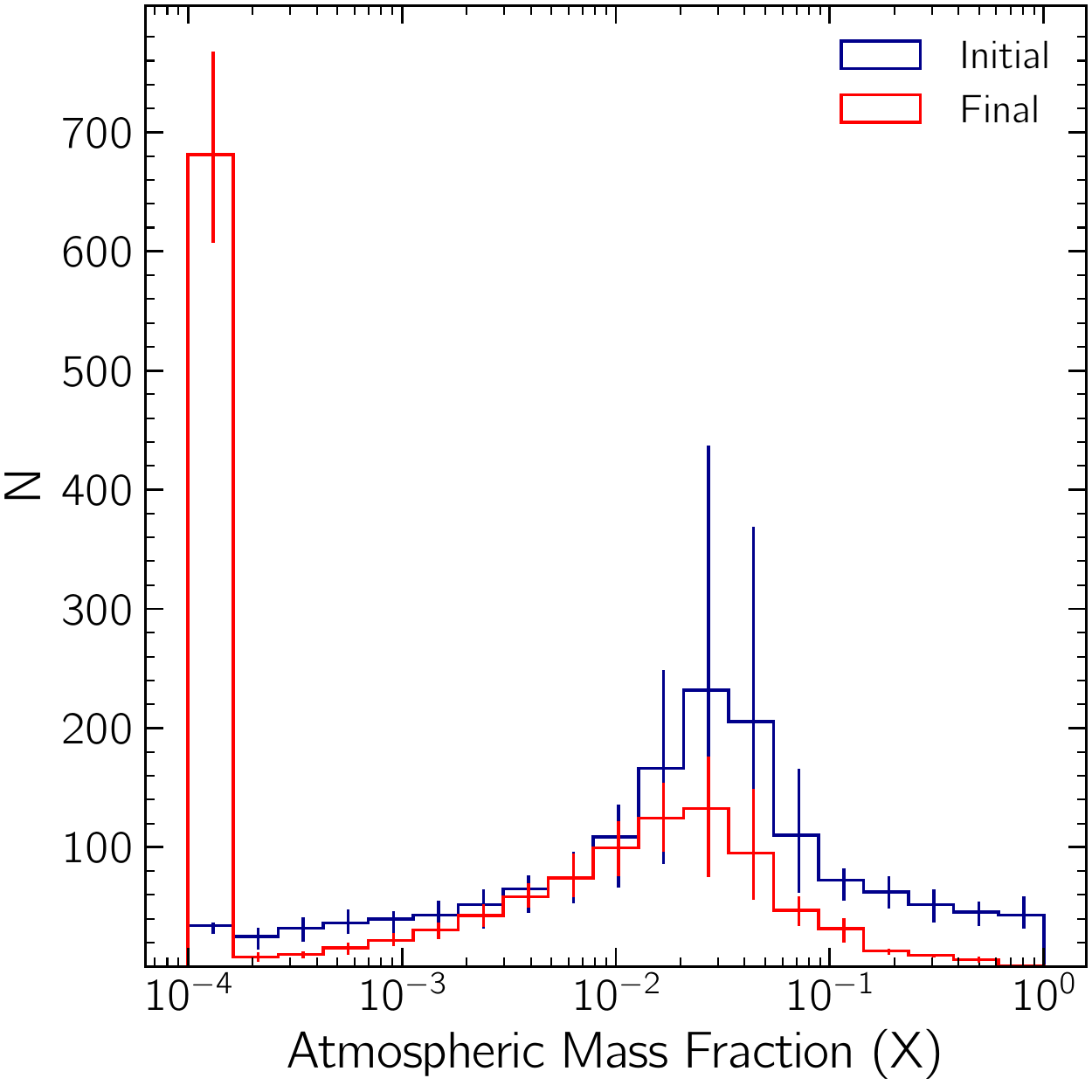}
    \caption{Evolution of atmospheric mass fractions from \textsc{model III}. Initial mass fractions shown in blue come from the constrained distribution shown in Figure \ref{fig:Combined-models}, whilst final mass fractions have been evolved through 3~Gyrs of photoevaporation. Error bars come from resampling from MCMC chain. The final distribution shows a large spike at $X=10^{-4}$, which corresponds to planets which have been completely stripped of their H/He atmospheres.}
    \label{fig:AtmosphereEvolution} 
\end{figure}

The inferred atmospheric mass fraction distribution is strongly peaked at $\sim 4\%$. Unlike the previous results in \citet{Owen2017} who only put weak constraints on the initial mass fraction we are able to constrain a precise distribution. This is due to the fact that photoevaporation becomes ineffective at longer orbital periods. Hence, the larger observed sub-Neptunes at longer periods in the CKS sample provide upper limits on the atmospheric mass fraction distribution. This result is further evidence that fitting the planet population in the period-radius plane (as opposed to the 1D radius distribution) is crucial in order to reveal more information of the underlying demographics.

Furthermore, the atmospheric mass fraction distribution points heavily towards the fact that the majority of Kepler planets were not ``born terrestrial'' i.e. with negligible atmosphere. Figure \ref{fig:AtmosphereEvolution} demonstrates the evolution of atmospheric mass fractions according to \textsc{model III}, with initial and final populations represented in blue and red respectively. As found in \cite{WolfgangLopez2015}, the final distribution shows a peak at a few percent, as this corresponds to the population of sub-Neptunes, i.e. those with large extended H/He atmospheres that double their core radius. On the other hand, the large spike of evolved planets with an atmospheric mass fraction of $X=10^{-4}$ indicates that these planets have been stripped of their natal H/He atmosphere. As a result, these bare rocky cores form the large population of super-Earths that we observe today. In order to quantify this, we estimate the ratio between number of planets that were born rocky, and number of planets that evolved to become so. By integrating the initial and final distribution below $X=10^{-3}$, we find that $\sim 4$ times as many planets evolved to become super-Earths as a result of photoevaporation, rather than be born with negligible atmosphere. We note that our results do indicate that at least some of the super-Earth's formed without a significant H/He atmosphere; however it is clearly a sub-dominant mode. We therefore gain insight into the origin of super-Earths and sub-Neptunes and can state that they both predominantly evolved from the same initial population. These results are consistent with \cite{NeilRogers2020}, in which three sub-populations of small, close-in exoplanets are identified: sub-Neptunes, super-Earths that were photoevaporated and super-Earths that were born terrestrial. Although each of the sub-populations was identified with a separate mass function, as opposed to our method which used a common function for all cores, both inference problems find a best fit to the data arises with when all three populations are present. \citet{OMC19} found evidence that the population of born terrestrial planets was common around lower-mass or lower-metallicity stars and speculated that the transition between accreting or not accreting a large H/He atmosphere was related to the supply of solids to the inner disc. It is worth emphasising that core accretion models \citep[e.g.][]{LeeChiang2015} indicated a 1~M$_\oplus$ core will accrete a 1\% H/He atmosphere by mass under typical nebula conditions. Therefore, while we see these born terrestrial planets with radii and masses above Earth's today, they must have been significantly less massive at the time of gas disc dispersal and continued accretion after disc dispersal, akin to the standard model for terrestrial planet formation in the Solar System.

\subsection{Core Mass Distribution} \label{sec:DiscussCoreMass}

\begin{figure} 
	\includegraphics[width=\columnwidth]{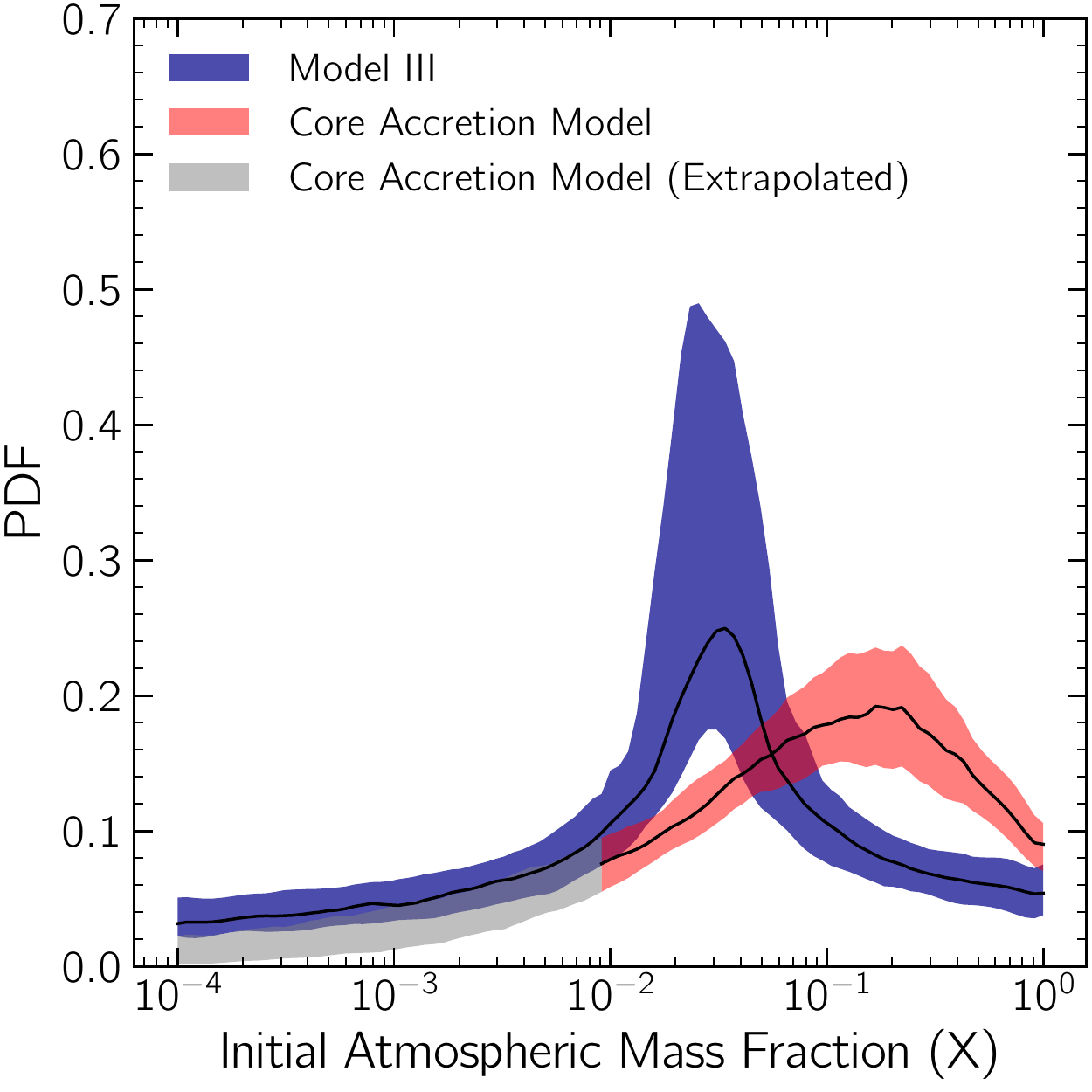}
    \caption{Comparison of inferred initial atmospheric mass fraction from \protect\textsc{model III} (shown in blue) with core accretion model from \protect\cite{LeeChiang2015}, adapted in \protect\cite{Jankovic2018} (Equation \ref{eq:coreacc}). Red region shows the predicted atmospheric mass accreted, assuming the core mass distribution found in \protect\textsc{model III}. Grey region represents accreted atmospheric mass fractions calculated by extrapolating our constrained core mass distribution to smaller cores masses ($<0.6M_\oplus$).}
    \label{fig:CoreAccretion} 
\end{figure}

\begin{figure*} 
	\includegraphics[width=2.1\columnwidth]{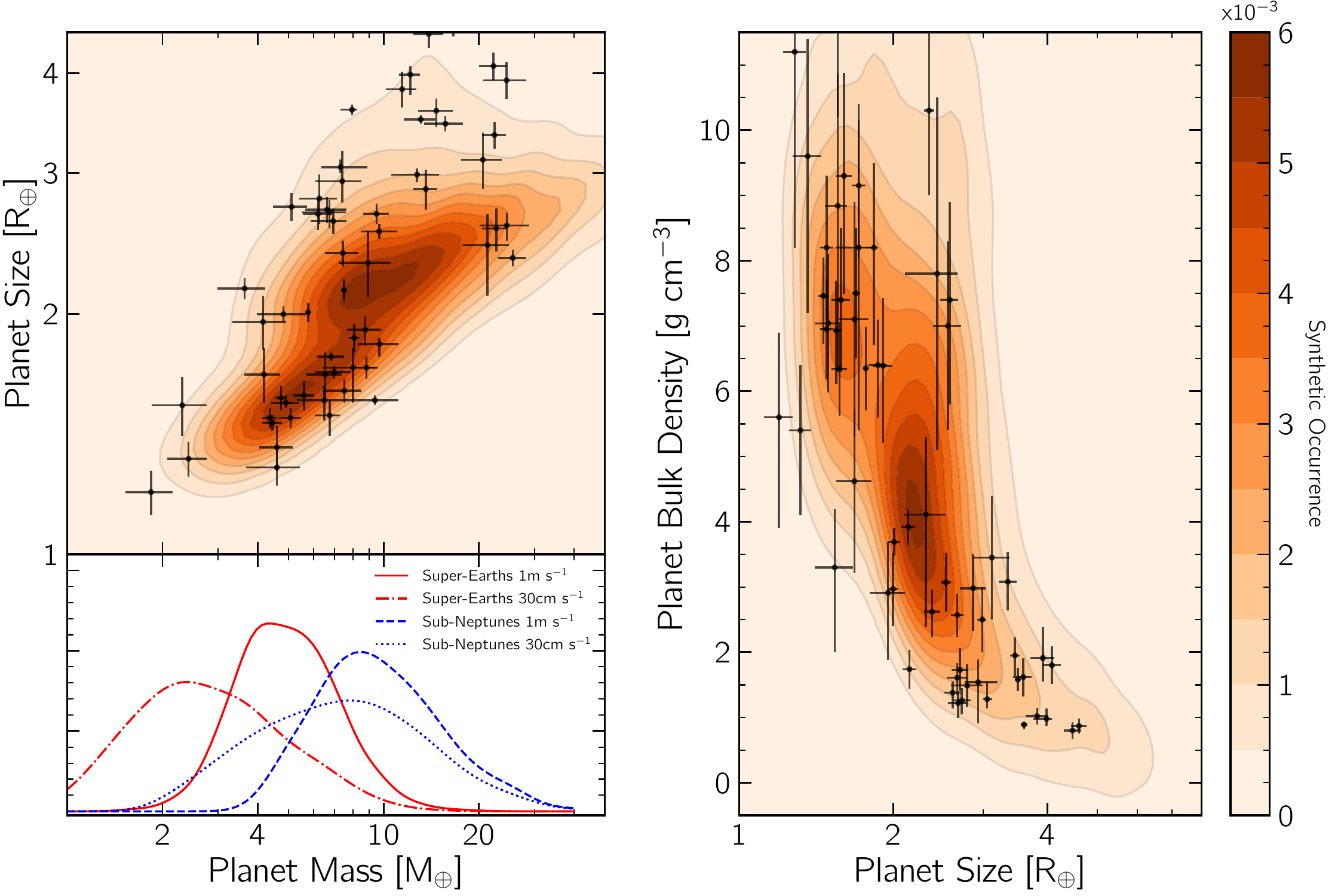}
    \caption{Left hand panel shows mass-radius plot for forward modelled planets from \textsc{model III} for which RV semi-amplitude is $>1\text{ms}^{-1}$. Whilst we have added $6\%$ fractional errors to planet radii and $20\%$ errors to planet masses, we have not included survey biases similar to those in the main inference problem. In addition, planet mass PDFs are also provided for super-Earths and Sub-Neptunes, defined by $R_p < 1.8 R_\oplus$ and $R_p > 1.8 R_\oplus$ respectively with minimum RV semi-amplitude of $1\text{ms}^{-1}$ (solid and dashed respectively) as well as $30\text{cms}^{-1}$ (dash-dotted and dotted respectively). Black points represent confirmed planets taken from the \textit{NASA Exoplanet Archive}, for which mass and radius measurement uncertainty is $<20\%$. In addition we do not include biases of surveys, hence this should be interpreted as an underlying mass-radius distribution.}
    \label{fig:MassRadius} 
\end{figure*}

The core mass distribution, shown in Figure \ref{fig:Combined-models}, indicates a strong peak at $\sim 4M_\oplus$, additionally confirming previous works which required a peaked distribution to fit data \citep[e.g.][]{Owen2017,Wu2019}. This feature implies that a common core mass and thus formation pathway is favoured in planetary systems. In addition, the steeper slope beyond $\sim 20 M_\oplus$ indicates that very few cores are produced beyond this range. This result confirms the predictions that cores growing to this size will begin to host self-gravitating atmospheres and undergo runaway gas accretion \citep{Pollack1996}. As these planets grow towards Jovian masses, they reach the ``thermal mass'' and may carve gaps in their protoplanetary discs, perhaps slowing accretion. This idea was put forward by \citet{Wu2019} as to the origin of the peak, which was strengthened by the fact they found this peak mass became smaller at lower-stellar mass. This trend is something that should be investigated in our inference framework, but is not possible with the narrow range of stellar masses in the CKS dataset and the use of non-parametric functions for the core-mass distribution. Nevertheless, we do require a non-zero fraction of cores $\gtrsim 20~$M$_\oplus$ to explain the data. This appears consistent with recent {\it NGTS} and {\it TESS} detections that have discovered a number of 30-40~M$_\oplus$ planets residing in the desert \citep{West2019,Jenkins2019,Armstrong2020}. While these planets are certainly consistent with photoevaporation, how they formed and why they did not accrete a massive H/He envelope remains a mystery. 

One benefit of the adopted EUV/X-ray photoevaporation evolution model is that it is capable of providing core mass, final atmospheric mass and photospheric radius for an ensemble of planets. Thus, one can produce a synthetic mass-radius diagram as shown in the left-hand panel of Figure \ref{fig:MassRadius}. Here we have added $6\%$ fractional errors to planet radii and $20\%$ errors to planet masses. In addition we have limited planets to have RV semi-amplitude of $>1\text{ms}^{-1}$, which represents current detection limits. Note however, that we have not included survey biases (as performed in our inference model) as these are currently not quantifiable for mass measurements. As a result, the mass-radius plot we provide should be interpreted as an underlying distribution with added noise. As observed in multiple works \citep[e.g.][]{Weiss2014,Marcy2014,HaddenLithwick2014,Dressing2015,JontofHutter2016,Dorn2019}, we see a large population on a single composition line (i.e. super-Earths with terrestrial cores) and another population at a larger radius for a similar mass. This latter group corresponds to the sub-Neptune population, harbouring extended H/He atmospheres that approximately double their observed radius. As a result, the mass distribution, as observed with RV or TTV surveys, for super-Earths and sub-Neptunes will be different. This is demonstrated in the left-hand panel of Figure \ref{fig:MassRadius}, in which the relative occurrence for super-Earths ($R_p < 1.8 R_\oplus$) and sub-Neptunes ($R_p < 1.8 R_\oplus$) is separated and plotted as PDFs, with sub-Neptunes typically occurring at larger masses. We also plot the same PDFs for a minimum RV semi-amplitude of $30\text{cms}^{-1}$ which represents the precision on next-generation RV instruments. We see that by improving this minimum value allows smaller planets to be detected and thus both distributions extend further to lower mass. In addition we also take planets from the \textit{NASA Exoplanet Archive}\footnote{Accessed on 04/04/2020.} with mass and radius measurement uncertainty $<20\%$ and plot them on top of our expected occurrence. In the right-hand panel of Figure \ref{fig:MassRadius}, we show the bulk density for the same population of modelled planets as a function of planet size. A clear bimodality is observed, with super-Earths centred at $\sim 6\text{ g cm}^{-3}$, owing to their terrestrial composition, whilst sub-Neptunes lie at a lower bulk-density $\sim 2\text{ g cm}^{-3}$ as a small increase in H/He atmosphere increases the planet's photospheric radius non-linearly.

An interesting question one can pose is, given the inferred core mass distribution, how much H/He should the cores accrete whilst immersed in a protoplanetary disc? Taking this further, we can determine if the constrained core mass distribution can predict the constrained initial atmospheric mass fraction distribution. To do so, we employ a core accretion scaling relation from \cite{LeeChiang2015} for a dust-free H/He atmosphere in order to calculate the accreted mass fraction for a given core mass\footnote{Although, as shown by \citet{Jankovic2018} choosing this specific model over others makes little difference}. As used in \cite{Jankovic2018}, the scaling relation is adapted for varying gas surface density \citep{Lee2018,FungLee2018}:
\begin{equation} \label{eq:coreacc}
    \begin{split}
    X(t) = & \; 0.18 \; \bigg( \frac{t}{1\text{ Myr}} \bigg)^{0.4} \; \bigg( \frac{0.02}{Z} \bigg)^{0.4}
    \; \bigg( \frac{\mu}{2.37} \bigg)^{3.3} \\
    & \times \; \bigg( \frac{M_\text{core}}{5M_\oplus} \bigg)^{1.6} \; \bigg( \frac{1600\text{ K}}{T_\text{rcb}} \bigg)^{1.9} \; \bigg( \frac{f_\Sigma}{0.1} \bigg)^{0.12}
    \end{split}
\end{equation}{}
where $Z$ is atmospheric metallicity, $\mu$ is mean molecular weight, $T_\text{rcb}$ is the temperature at the radiative-convective boundary inside the atmosphere and $f_\Sigma$ is the ratio of gas surface density to that of the minimum mass solar nebula \citep{Hayashi1981}. Figure \ref{fig:CoreAccretion} shows the inferred initial atmospheric mass fraction distribution (from \textsc{model III}) in blue, whilst the accreted mass fraction (according to Equation \ref{eq:coreacc}) in red. In order to produce a full range of atmospheric masses, the constrained core mass distribution is extrapolated below the range used in our inference model - this untrustworthy region of parameter space is shown in grey. We see clearly that the atmospheric mass predicted by core accretion is $\sim5$ times larger than our constrained distribution.

This discrepancy has been found in other previous works. In \cite{Jankovic2018}, magneto-rotational instability (MRI) simulations were performed to reveal that the masses of cores forming in the inner disc would result in atmospheres that were too large to be photoevaporated and thus evolve into the typical size of planet we observe today. Possible mechanisms to resolve this tension include forming the planets at the very end of the disc life-time \citep{Ikomi2012,LeeChiang2016}, or alternatively improving the accuracy of gas accretion models to include the effects of giant mergers \cite{Liu2015,Inamdar2016} which would result in potentially significant atmospheric mass-loss and the production of heat which would take typically kyrs to disperse. The inclusion of 3D simulations has additionally shown that recycling of high-entropy gas during the accretion phase can act to reduce the final atmospheric mass of the planet \citep{Ormel2015,Fung2015,Cimerman2017,AliDib2020,Chen2020}.

Finally, a different approach is the hypothesised `boil-off' mechanism  \citep{Owen2016} in which, during protoplanetary disc dispersal, low-mass exoplanets may launch a Parker wind \citep{Parker1958} due to the quick reduction in confining gas pressure. This rapid mass-loss mechanism may account for the disagreement between our constrained atmospheric mass fraction distribution and that provided by core accretion theory.

\subsection{Core Composition}

\begin{figure} 
	\includegraphics[width=\columnwidth]{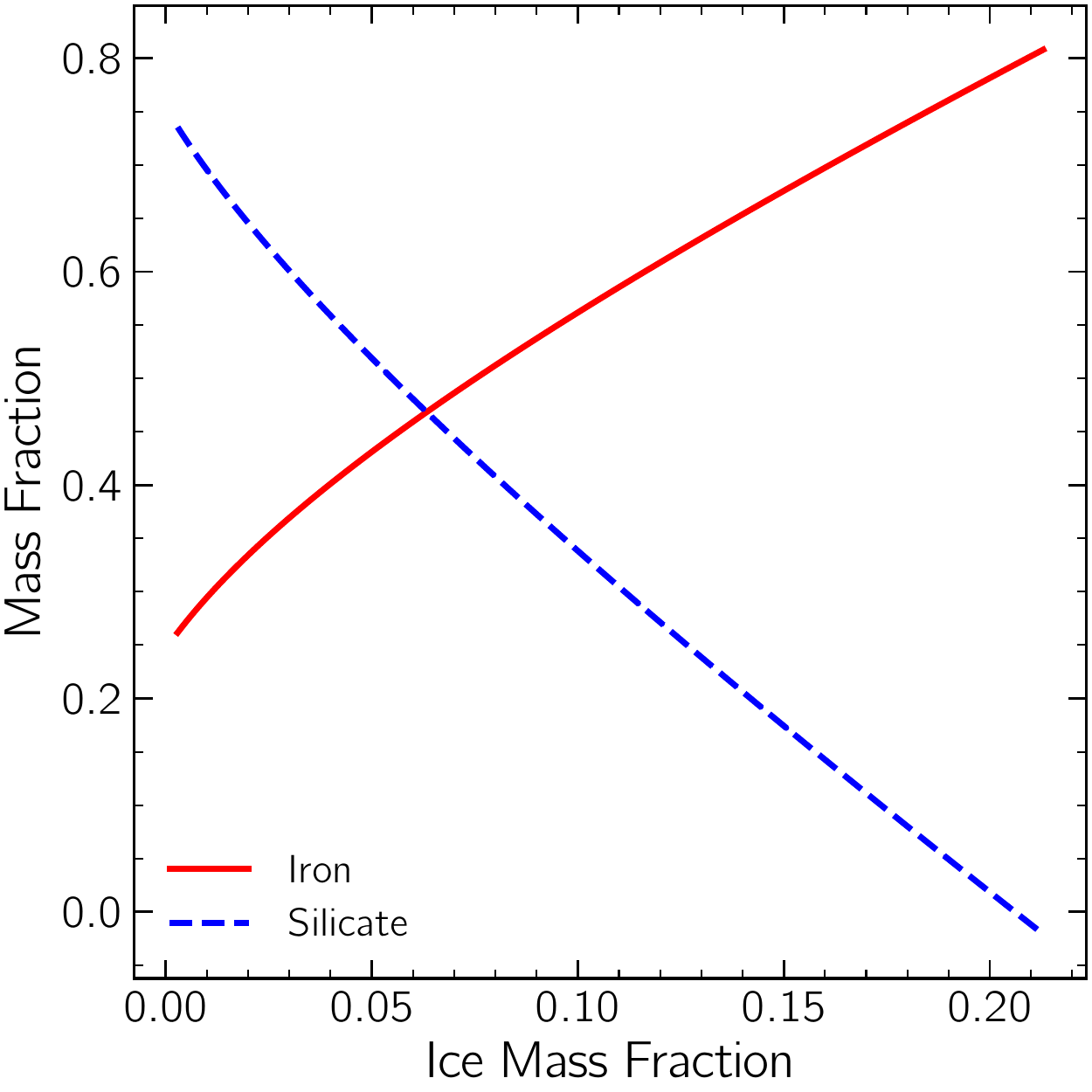}
    \caption{Mass fractions of iron and silicate as a function of ice mass fraction for a $4M_\oplus$ planet according to core interior models from \protect\cite{Zeng2008}. We fix the core radius to be $\sim 1.63R_\oplus$ in this calculation as this requires that an ice-free core has a silicate-to-iron composition ratio of $\sim$ 3:1 and hence aligns with our constrained distributions.}
    \label{fig:IceFractions} 
\end{figure}

As with all models presented in this work, the inferred mean core composition is found in a silicate/iron dominated regime. Taking the value from \textsc{model III}, we find a mean value of $\tilde{\rho} = 0.26^{+0.08}_{-0.09}$, which is equal to a bulk core density for a $1M_\oplus$ core of $\rho = 5.10^{+0.39}_{-0.40} \text{ g cm}^{-3}$. Interpreting this composition using the \cite{Fortney2007} mass-radius relations, suggests that cores of this nature would have a silicate-to-iron ratio of $\sim$3:1, which is consistent with that of Earth and confirms findings of previous works \citep[e.g.][]{Owen2017,Wu2019}. Additionally, the typical bulk core density of observed planets beneath the gap (and hence host negligible atmosphere) are consistent with this value \citep{Dressing2015,Dorn2019}. Taking this interpretation of the core composition further, it follows that in order to form cores dominated by silicate material and absent of ices, it is necessary to build them interior to the water-ice line \citep{Chatterjee2014,Jankovic2018}. A possible restriction in our adopted methodology is that our core composition analysis has implemented the \cite{Fortney2007} mass-radius relationships, which only allow two species to be present in a core at one time: either an ice-silicate mixture or a silicate-iron mixture. In reality, all three species are likely to coexist in cores and hence we use an exoplanet interior model from \cite{Zeng2008} to calculate the allowed composition fractions for a planet with our inferred bulk core density $\sim 5\text{g cm}^{-3}$. Figure \ref{fig:IceFractions} shows the iron and silicate mass fractions as a function of ice mass fraction for a $4M_\oplus$ planet with a $\sim 1.63R_\oplus$ core radius - not only does this combination ensure matching bulk core densities with our inferred distributions, but also is a typical mass and radius for a super-Earth. We see that in order to increase the ice mass fraction by a small amount, the iron content must increase dramatically. Furthermore, if we wish to match ice mass fractions predicted by formation models beyond the water-ice line, the iron mass fraction must rise to unphysical levels and remove the vast majority of the silicate mass. We thus conclude that, although our adopted composition interpretation is limited to two species, only a negligible ice content would be introduced with more sophisticated interior models.  Although this points towards the \textit{in-situ} formation pathways, additional physics such as the presence of planetary magnetic fields \citep{OwenAdams2019} can act to suppress atmospheric mass-loss, leading to a lower inferred core density and hence ice-rich cores. This would, however, result in an inconsistency between inferred planet masses and those observed with RV or TTV surveys.

\subsection{Core Composition Spread}
In fitting a Gaussian function to core compositions, \textsc{model III} not only provides a mean-value, but also a width to the core composition distribution. As shown in Figure \ref{fig:modelIII-posteriors}, we can only place a $2\sigma$ upper limit on this width to be $\sigma_{\tilde{\rho}} < 0.16$, which is evidently consistent with zero. This composition spread corresponds to $2\sigma$ upper limit in the variation of the density of an Earth-mass core of $<1.5$~g~cm$^{-3}$.  It was shown in \cite{Owen2017} that for a decrease in bulk core density for constant mass, a planetary core expands to a larger radius. As a result, a larger atmosphere can be stripped and thus the location of the radius gap moves to a larger value. To produce a clean gap i.e. one with a greater sparsity of planets, such as that found in \cite{VanEylen2018}, the spread in bulk core densities must be very narrow. As this has now been quantified in \textsc{model III}, it suggests that, despite different host stars and stellar neighbourhoods, small, close-in exoplanet cores all attain a very similar core density. Whilst there exist strong degeneracies in core composition, which allow different fractions to equate to the same bulk core density, it still raises the question of why such fine-tuning of bulk density would arise in a stochastic core assembly process \citep{Michel2020}.

\begin{figure} 
	\includegraphics[width=\columnwidth]{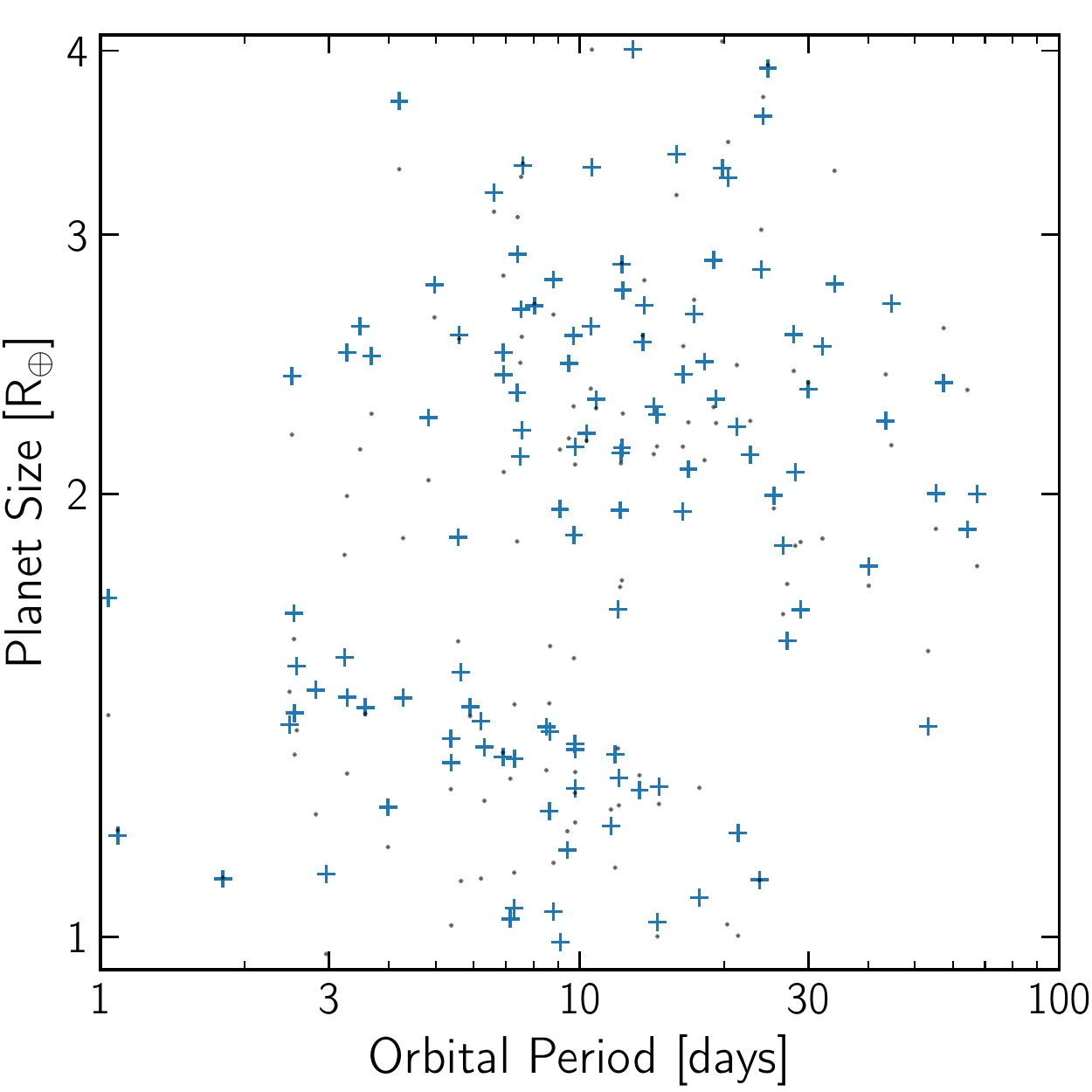}
    \caption{ {\correction As a comparison to the asteroseismic survey of \protect\cite{VanEylen2018}, here we show 117 forward modelled planets randomly selected from \textsc{model III}. Blue crosses are planets where no radii errors are added, hence we produce a clean gap. Black dots represents the same planets with typical CKS errors added instead, which can shift some planets into the gap. Hence, despite using a data set with larger radii uncertainties, we still predict a clean gap in the underling planet distribution.} }
    \label{fig:CleanGap} 
\end{figure}

{\correction It is worth noting that whilst the cleaner gap found in the asteroseismic sample of \cite{VanEylen2018} is more accurate and therefore desirable \citep{Petigura2020}, the survey is currently not large enough to yield meaningful results if included in this forward modelling methodology. In addition, in order for the model to be computationally efficient, a completeness map is required in order to calculate synthetic planet detections, which is not available for the \cite{VanEylen2018} sample. 
To check whether our model can produce a clean gap in the absence of radius errors, Figure \ref{fig:CleanGap} shows a random draw of 117 planets (the same as the \citealt{VanEylen2018} sample) forward modelled using \textsc{model III} with and without radius errors included. This shows that the underlying distribution of planets produces a clean radius gap, and it is the uncertainty in planetary radii that pushes planets into the valley.}

\subsection{Planet Distribution Beyond 100 day Orbital Period}

\begin{figure} 
	\includegraphics[width=\columnwidth]{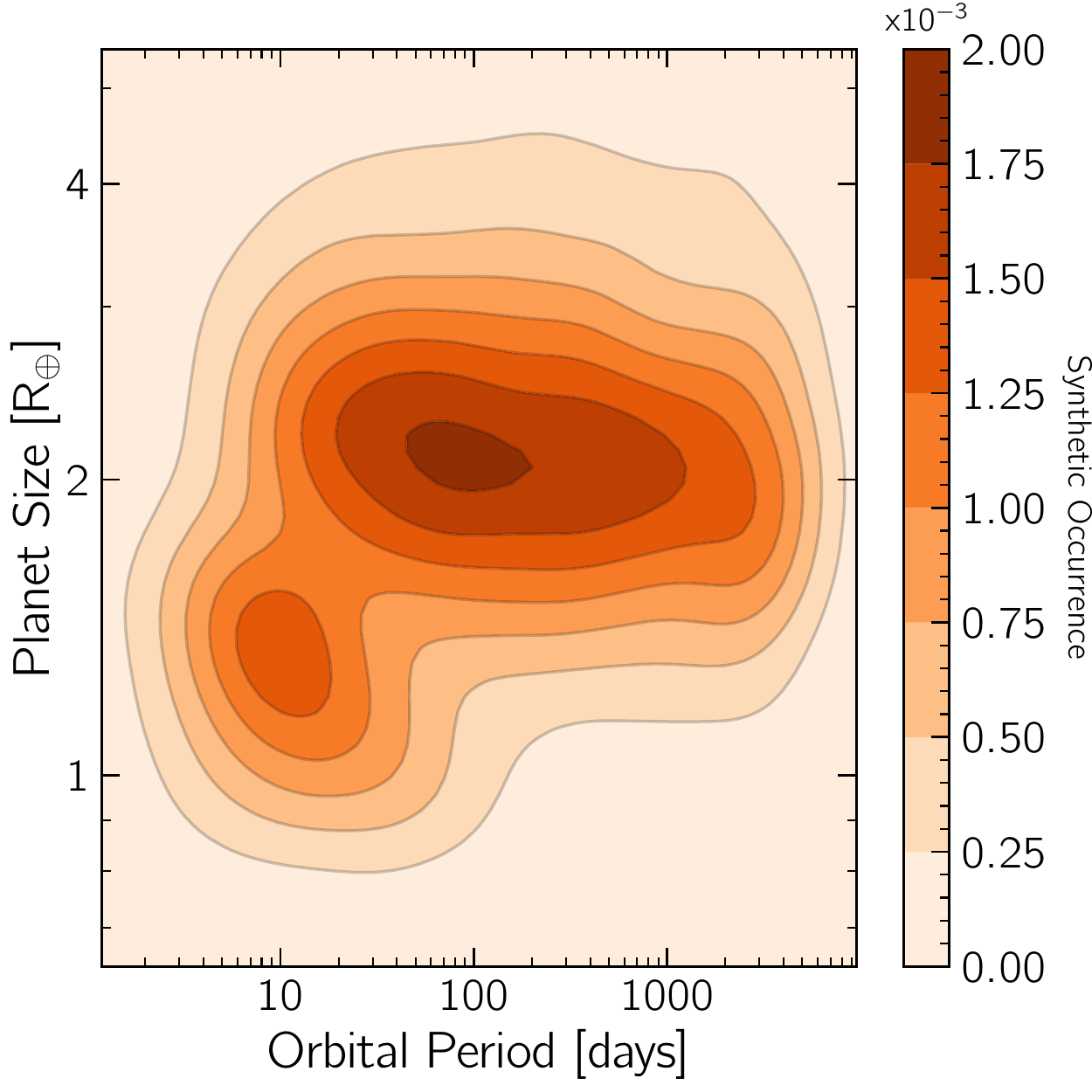}
    \caption{Planet size distribution as a function of orbital period for best fit \textsc{model III} parameters under the assumption that the inferred period distribution (Equation \ref{eq:period}) extends beyond 100 days.}
    \label{fig:LongPeriods} 
\end{figure}

Whilst current detection limits prevent surveys probing exoplanet distributions at larger orbital periods, one can postulate as to whether the distribution we observe today extends beyond 100 days. As shown in Figure \ref{fig:LongPeriods}, extrapolating the period distribution used in this work (Equation \ref{eq:period}) and modelling planets according to \textsc{model III} demonstrates that the sub-Neptune population would in-fact be far greater in number than that of the super-Earths. This is because planets at greater orbital separations are less prone to atmospheric mass-loss and therefore retain a few percent mass atmosphere. If, on the other hand, we were to observe a distribution in which the relative occurrence of super-Earths and sub-Neptunes remains approximately equal beyond 100 days, it would suggest that the number of planets that were born terrestrial might increase with orbital period.  

\subsection{Changing the Evolution Model}

\begin{figure} 
	\includegraphics[width=\columnwidth]{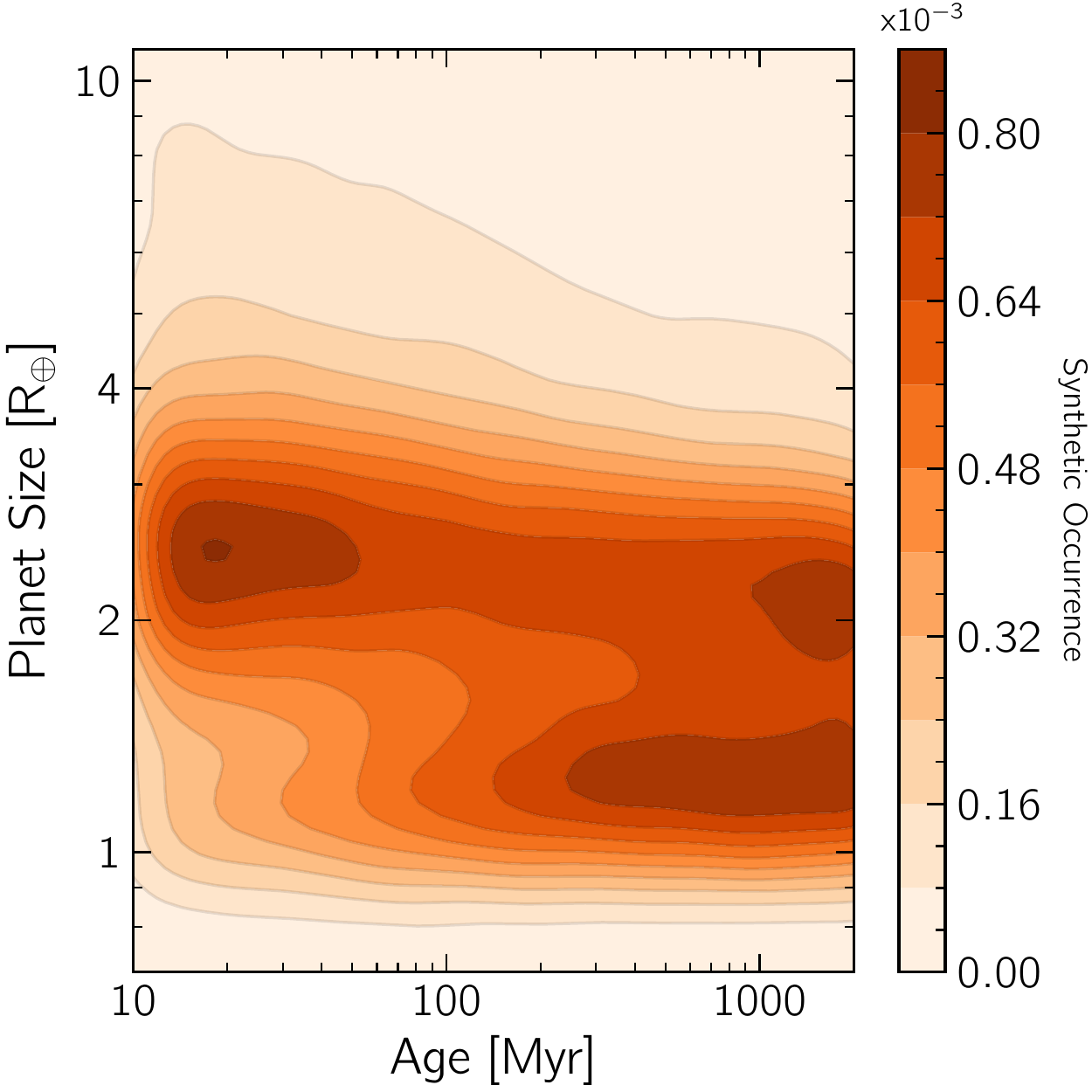}
    \caption{Planet size distribution as a function of time (or similarly host star age). The bimodality can be seen to emerge after 100 Myrs, typically when the star begins to spin down and the XUV luminosity drops off precipitously. When compared to core-powered mass-loss (i.e. Figure 10 in \protect\cite{Gupta2020}) we see the timescale for radius gap emergence is of order Gyrs.}
    \label{fig:RadiusEvolution} 
\end{figure}

As discussed in Section \ref{sec:Method}, all results presented in this work hinge on the EUV/X-ray photoevaporation model and accurate photoevaporation rates, which forms a strong prior on on all constrained quantities. Other evolution models, such as core-powered mass-loss, are similarly capable of reproducing the observed data \citep{Gupta2019,Gupta2020}. Whilst both models have similar dependencies and hence inferred quantities (e.g. core composition), they differ in some regards, particularly in planet radius evolution. As the majority of the mass-loss in photoevaporation occurs during the first few 100  Myrs before the host star begins to spin down, the radius gap is predominantly formed at the end of this stage (see Figure \ref{fig:RadiusEvolution}). Core-powered mass-loss on the other hand invokes the core's accretion luminosity to provide the energy source required to induce atmospheric mass-loss. As the time scales for this energy-transfer are far greater than that of EUV/X-ray photoevaporation, the radius gap takes far longer to form, typically of order Gyrs. This difference in evolution my lead to different inferred demographics, especially initial atmospheric mass fraction. We therefore suggest that comparisons of inferred quantities from both models can act as further tests between the two mechanisms. As a very basic statistic, we measure the occurrence ratio of super-Earths to sub-Neptunes defined by number of planets with $R_p \leq 1.8R_\oplus$ and $R_p > 1.8R_\oplus$ respectively. We find ratio values of $0.77\pm0.08$ for stellar ages $t_* < 1\text{ Gyr}$ and $0.95\pm0.08$ for $t_* > 1\text{ Gyr}$. This indicates that, while the majority of planets cross the gap on 100~Myr time-scales, there are a minority that do so at later times (as can be seen in Figure~\ref{fig:RadiusEvolution}). Works are beginning to constrain the time-dependence of the exoplanet population (such as \citealt{Berger2020}) but, a single cut at a specific time is particularly sensitive to the underlying age distribution either side of the cut. However, trying to constrain the evolution of the ratio of super-Earths to sub-Neptunes as a function of time will provide a constraint on the driving mass-loss model.   

\subsection{Improving the Model}
In general, this work has been a proof of concept that observationally unobtainable demographic quantities can be inferred from pre-existing data by utilising an evolutionary model. One key limitation is the amount of data included. As new transit surveys such as \textit{TESS} \citep{RICKER2014} begin to reveal a myriad of small, close-in exoplanets, the inclusion of the such data will only increase the constraining power of the model. Despite what we have shown is possible, the limited CKS sample ($\sim 900$ planets) meant that we limited our non-parametric Bernstein polynomial approximations of core mass and initial atmospheric mass fraction distributions to $5^\text{th}$-order. {\correction To confirm a $5^\text{th}$-order expansion is sufficient to characterise the distributions, we performed a repeat of \textsc{model III} to convergence with $8^\text{th}$-order Bernstein polynomials. This test yielded results that were consistent with the ones shown here. We therefore conclude that with the current limited data, we are unable to extract more detailed features in the distributions of interest.} In addition to more transit data, mass measurements may be included such as from RV or TTVs, again improving the constraining ability of the model. We do note however, that the observational biases of any new data must be quantifiable in order to include it in our inference model.

In the presence of more data, more complex trends can be explored. One such trend that was probed in \cite{Wu2019} was the dependency of core mass with stellar mass i.e. do larger stars host larger planets? We therefore aim to determine whether the peak of the inferred core mass distribution increases for larger host stars. Furthermore, one might expect the core mass and initial atmospheric mass fraction distributions to be correlated due to the fact that larger cores are able to accrete larger atmospheres. Furthermore, more data could be used to determine dependencies of core mass or initial atmospheric mass fraction distributions with stellar mass or orbital period, the latter of which might aid in placing constraints on core formation pathways, atmospheric accretion rates or `boil-off' style mass-loss mechanisms. Stellar mass trends on the other hand will be pivotal in comparing photoevaporation with core-powered mass-loss. A key prediction of photoevaporation is that, at lower stellar mass, stars produce a higher {\it relative} EUV/X-ray flux, and are hence capable of stripping larger atmospheres for a fixed equilibrium temperature. On the other hand, core-powered mass-loss will have no such dependency on stellar mass as the energy source comes from within the planet's core. Extracting the presence, or lack of, these trends will be a crucial step in determining which mechanism is the dominant driver of planet evolution.

A final note is that the success of this inference model is pivotal on the fact that the analytic model from \cite{Owen2017} is capable of rapidly calculating planetary evolution due to photoevaporation. The down-side of this however, is that this is an approximation of full numerical simulations, such as those from \cite{OwenJackson2012,Owen2013}. As a result, accurately modelling aspects such as mass-loss efficiency, atmospheric opacity and envelope equation of state is limited. Further refinements to such analytic models is thus also worth investing time in. 

\section{Conclusion}
In this work we have used an evolutionary model for EUV/X-ray photoevaporation and the California Kepler Survey (CKS) data to infer the core mass distribution, the initial atmospheric mass fraction distribution and the core composition distribution for small close-in exoplanets. This is achieved by invoking the photoevaporation model to evolve populations of exoplanets and then synthetically observe them with the observational biases and noise of the \textit{Kepler Space Telescope}. By then comparing the modelled detections with the real CKS data, we infer which underlying planet demographics are required to match the model with the data. Our main conclusions are as follows:
\begin{itemize}
    \setlength\itemsep{1em}
    
    \item The core mass distribution is peaked at $\sim 4M_\oplus$, with a steep decline of occurrence towards higher masses. This points towards a singular formation pathway of planetary cores, preferentially produced at a few earth masses.
    
    \item The core composition distribution is centred at a bulk core density for an Earth mass core of $\rho_{1~{\rm M}\oplus} \approx 5.1 \text{ g cm}^{-3}$. This corresponds to 4:1 to 3:1 silicate iron composition ratio, which is consistent with Earth and heavily favours forming cores interior to the water ice-line.
    
    \item The core composition distribution has an extremely narrow width, which is consistent with zero. This suggests a fine tuning of bulk core density, despite difference in host stars and stellar neighbourhoods.
    
    \item The initial atmospheric mass fraction distribution is strongly peaked at $\sim 4\%$. Evolving this forward in time shows that approximately four times as many planets evolved to become stripped rocky cores, than those that were born rocky with no extended H/He atmosphere.
    
    \item Core accretion models over-predict the initial atmospheric masses of small exoplanets when compared to that which photoevaporation can strip. To resolve this tension, more sophisticated simulations of core accretion are needed. Alternatively, additional mass-loss mechanisms such as late-stage giant mergers or `boil-off' phases may further reduce the atmospheric mass and hence bring these models into agreement.
    
\end{itemize}

These conclusions are all dependant on the fact that we assume the main driver of small, close-in exoplanet evolution is EUV/X-ray photoevaporation. Changing the adopted evolution model to alternative theories, such as core-powered mass-loss may result in differing constrained distributions, which may provide tests between the models. In particular, determining trends with stellar mass may provide the best differentiation between the two mechanisms.

\section*{Acknowledgements}
We thank Subu Mohanty for useful discussions. {\correction We kindly thank the anonymous reviewer} as well as Emma Chapman, Erik Petigura, Yvonne Unruh and Yanqin Wu for comments that helped improve the paper. We are grateful to the CKS team for making their results public, which allowed this work to be performed. JGR is supported by a 2017 Royal Society Grant for Research Fellows and a 2019 Warner Prize. JEO is supported by a Royal Society University Research Fellowship. This work was supported by a 2020 Royal Society Enchantment Award and this project has received funding from the European Research Council (ERC) under the European Union's Horizon 2020 research and innovation programme (Grant agreement No. 853022, PEVAP). This work was performed using the DiRAC Data Intensive service at Leicester, operated by the University
of Leicester IT Services, which forms part of the STFC
DiRAC HPC Facility (www.dirac.ac.uk). The equipment
was funded by BEIS capital funding via STFC capital grants
ST/K000373/1 and ST/R002363/1 and STFC DiRAC Operations grant ST/R001014/1. DiRAC is part of the National e-Infrastructure.

\section*{Data Availability}
The MCMC chains underlying this article are publicly available on Zenodo, at \href{https://doi.org/10.5281/zenodo.4554210}{10.5281/zenodo.4554210}.

\appendix

\section{Bernstein Polynomials} \label{app:Bernstein}
Bernstein polynomials are effective in approximating well-behaved functions. They are formed as an $n^{th}$ order expansion of Bernstein basis polynomials:
\begin{equation} \label{eq:A1}
    B_n(x) = \sum^n_{\nu=0} \beta_{\nu,n} \, b_{\nu,n}(x),
\end{equation}
where $\beta_\nu$ are the Bernstein coefficients and the basis functions $b_{\nu,n}(x)$ are given by:
\begin{equation} \label{eq:A2}
    b_{\nu,n}(x) = \binom n \nu \, x^\nu \, (1-x)^{n-\nu} \,.
\end{equation}
For a given function $f(x)$ to be approximated, the Bernstein coefficients are calculated by:
\begin{equation} \label{eq:A3}
    \beta_{\nu,n} = f \bigg ( \frac{\nu}{n} \bigg) , 
\end{equation}
hence, combing equations \ref{eq:A1}, \ref{eq:A2}, \ref{eq:A3} leads to an approximation of $f(x)$ to $n^{th}$ order:
\begin{equation} \label{eq:Bernstein}
    B_n(f)(x) = \sum^n_{\nu=0} f \bigg ( \frac{\nu}{n} \bigg) \, \binom n \nu \, x^\nu \, (1-x)^{n-\nu}.
\end{equation}

In this work, Bernstein polynomials are used to model the initial atmospheric mass fraction distribution $f(X_\text{init})$ and core mass distribution $f(M_\text{core})$ in the inference model. For the latter, the Bernstein coefficients simply control the PDF of $\log M_\text{core}$ in the chosen domain between $0.6 M_\oplus$ and $100 M_\oplus$. The lower limit was chosen such that a pure ice composition would be at the current detection limit. The upper limit on the other hand was required to be sufficiently large to avoid boundary issues with the Bernstein polynomials. Whilst the coefficients for core mass control a PDF, the coefficients for $\log X^\text{init}_\text{atm}$ control an inverse CDF. On other words, we construct a Bernstein polynomial function in the domain $[0,1]$ and range $[\log X_\text{min}, \log X_\text{max}]$. By drawing a random number $x \in [0,1]$, we can therefore read off an initial atmospheric mass fraction according to the desired distribution. By adapting \ref{eq:Bernstein}, we can approximate our desired inverse CDF $C(x)$ as:
\begin{equation} 
    \begin{split}
        C(x) \approx  & \bigg\{ \frac{y_\text{max}-y_\text{min}}{B(1) - B(0)}  \\ & \times \sum^n_{\nu=0} C \bigg ( \frac{\nu}{n} \bigg) \, \binom n \nu \, x^\nu \, (1-x)^{n-\nu} \bigg\} + y_\text{min}
    \end{split}
\end{equation}
where $B(0)$ and $B(1)$ are equation \ref{eq:Bernstein} evaluated at $x=0,1$ for the Bernstein coefficients $\beta_{\nu,n} = C(\nu / n)$ to be determined in the inference problem. The choice of PDF vs. CDF between $f(X_\text{init})$ and $f(M_\text{core})$ came about due to convergence issues of MCMC chains with both distributions controlled by a CDF or both with a PDF. It was found that the core mass and initial atmospheric mass fraction distributions were effectively constrained if they were fit with a PDF and CDF respectively.

\section{Fitting the Orbital Period Distribution} \label{sec:PeriodAppendix}
The orbital period distribution is found by fitting the CKS data with a separate inhomogeneous point Poisson process, i.e. the same underlying statistical model used for the main inference problem of this paper. We approximate the underlying distribution for orbital period as a smooth broken power law.
\begin{equation} \label{eq:APperiod}
    \frac{\mathrm{d}N}{\mathrm{d}P} \propto \frac{1}{\big(\frac{P}{P_0}\big)^{-k_1} + \big(\frac{P}{P_0}\big)^{-k_2}},
\end{equation}
Once normalised as a PDF, we bias this function with a 1D integral of the completeness map of the CKS data set (i.e. integrate middle panel of Figure \ref{fig:completeness} along the $R$-direction). In this way, we calculate the probability of detecting a planet for a given orbital period, which we label $\lambda(P)$. As with our likelihood for the main inference problem (Equation \ref{eq:likelihood}) we construct a likelihood from an inhomogeneous point Poisson process, now in 1D,
\begin{equation} \label{eq:Periodlikelihood}
\begin{split}
L(\pi_\text{CKS}) & = P(N_\text{CKS}) \cdot P(\pi_\text{CKS}) \\
 & = e^{-\Lambda} \, \frac{\Lambda^{N_\text{CKS}}}{N_\text{CKS}!} \; \prod_i^{N_\text{CKS}} \frac{\lambda(P_i)}{\Lambda},
\end{split}
\end{equation}
where,
\begin{equation} \label{eq:PeriodLambda}
    \Lambda = N_* \cdot f_* \cdot \int \lambda(P) \: \mathrm{d} P ,
\end{equation}
is the total number of planets expected to be detected and $\pi_\text{CKS}$ is the population of CKS planets (see Section \ref{sec:Likelihood} for more details). This likelihood is given to the affine invariant MCMC of \cite{GoodmanWeare2010} from the \textsc{emcee} package \citep{emcee} with 500 walkers in order to infer $P_0$, $k_1$ and $k_2$ from Equation \ref{eq:APperiod}.

\section{Fitting the Host Stellar Mass Distribution} \label{sec:SmassAppendix}
Similar to the orbital period distribution, we fit the stellar mass distribution of \textit{Kepler} planets using an inhomogeneous point Poisson process. We do not, however, quantify the selection biases and instead choose to simply infer the underlying distribution as Gaussian function. In order to forward model the stellar mass errors, we take the typical fractional uncertainties from the CKS catalogue (typically $\sim$ 4\%) and add them to a population of stellar masses drawn from a Gaussian function. We then calculate a PDF from this new distribution of noisy stellar masses $\lambda(M_*)$ and, as with Equation \ref{eq:Periodlikelihood}, measure a likelihood by taking  the value of this PDF for each of the CKS stellar mass measurements. We provide this likelihood function to \textsc{emcee} \citep{emcee} in order to infer the mean and standard deviation of the underlying stellar mass distribution from the CKS catalogue.
\label{sec:MstarFitting}





\bibliographystyle{mnras}
\bibliography{references} 

\appendix

\bsp	
\label{lastpage}
\end{document}